\documentclass[fleqn,10pt]{wlscirep}
\usepackage[utf8]{inputenc}
\usepackage[T1]{fontenc}
\usepackage{multirow}
\usepackage{comment}
\usepackage{appendix}
\usepackage{booktabs}   % For professional quality tables
\usepackage{tabularx}   % For tables that span the text width
\usepackage{caption}    % For better control over captions
\usepackage[left]{lineno}
\usepackage{hyperref}
\usepackage{xr-hyper} % For cross-referencing with external files

\title{The changing surface of the world's roads}

\author[1,*]{Sukanya Randhawa}
\author[2]{Guntaj Randhawa}
\author[1,2]{Clemens Langer}
\author[1,2]{Francis Andorful}
\author[1]{Benjamin Herfort}
\author[2]{Daniel Kwakye}
\author[1]{Omer Olchik}
\author[1,2,3]{Sven Lautenbach}
\author[1,2,3]{Alexander Zipf}
\affil[1]{Heidelberg Institute of Geoinformation Technology (HeiGIT), 
Berliner Str. 45 (Mathematikon), 69120 Heidelberg, Germany}

\affil[2]{GIScience Chair, Institute of Geography, Heidelberg University, 
Im Neuenheimer Feld 368, 69120 Heidelberg, Germany}

\affil[3]{Centre for the Environment, Heidelberg University, 
Im Neuenheimer Feld 130.1, 69120 Heidelberg, Germany}

\affil[*]{sukanya.randhawa@heigit.org}


\begin{abstract}

Resilient road infrastructure is a cornerstone of the UN Sustainable Development Goals. Yet a primary indicator of network functionality and resilience is critically lacking: a comprehensive global baseline of road surface information. Here, we overcome this gap by applying a deep learning framework to a global mosaic of Planetscope satellite imagery from 2020 and 2024. The result is the first global \textit{multi-temporal} dataset of road pavedness and width for 9.2 million km of critical arterial roads, achieving 95.5\% coverage where nearly half the network was previously unclassified.

This dataset reveals a powerful multi-scale geography of human development. At the planetary scale, we show that the \textit{rate of change} in pavedness is a robust proxy for a country’s development trajectory (correlation with HDI = 0.65). At the national scale, we quantify how unpaved roads constitute a fragile backbone for economic connectivity. We further synthesize our data into a global \textit{Humanitarian Passability Matrix} with direct implications for humanitarian logistics. At the local scale, case studies demonstrate the framework’s versatility: in Ghana, road quality disparities expose the spatial outcomes of governance; in Pakistan, the data identifies infrastructure vulnerabilities to inform climate resilience planning. Together, this work delivers both a foundational dataset and a multi-scale analytical framework for monitoring global infrastructure, from the dynamics of national development to the realities of local governance, climate adaptation, and equity. Unlike traditional proxies such as nighttime lights, which reflect economic activity, road surface data directly measures the physical infrastructure that underpins prosperity and resilience—at higher spatial resolution.

\end{abstract}

\begin{document}

\flushbottom
\maketitle
\thispagestyle{empty}

\section*{Introduction}

%The Introduction section, of referenced text\cite{Figueredo:2009dg} expands on the background of the work (some overlap with the Abstract is acceptable). The introduction should not include subheadings.

Road networks are the arteries of the global economy, essential for trade, social integration, and access to fundamental services such as healthcare and education\cite{worldbank_transport_2017,10.1093/jeea/jvab027}. The quality of this infrastructure, determined largely by its surface type, plays a pivotal role in determining transportation efficiency, disaster resilience, and progress towards the United Nations Sustainable Development Goals\cite{un_sdg9,Wenz_2020, african2014tracking}. The distinction between paved and unpaved roads shapes not only per capita income levels\cite{Queiroz1992Roads} but also a nation’s vulnerability to climate-related disruptions such as floods and extreme weather events\cite{calderon_roads_2015,GEBRESILASSE2023103048,Koks_2023}. Consequently, accurate, high-resolution and up-to-date global data on road surface conditions is not merely a logistical asset but a fundamental prerequisite for monitoring economic progress, targeting infrastructure investment, and assessing climate vulnerability.

While previous research has relied heavily on proxies such as nighttime light intensity to infer economic activity and human development\cite{doi:10.1126/science.aaf7894,yeh2020economic}, such radiance-based measures remain limited by their coarse spatial resolution and their indirect relationship to the underlying physical infrastructure that enables growth. In contrast, road surface characteristics offer a direct, high-resolution, and functionally grounded signal of economic investment, accessibility, and resilience—capturing dimensions of human development that nighttime lights cannot resolve.

Yet despite this critical need — and substantial progress made by large-scale geospatial AI - a comprehensive global baseline of road surface information remains elusive at the time of writing. Recent years have seen remarkable progress in deriving geospatial datasets of the built environment, driven by large-scale AI initiatives. Recent datasets from organizations like Meta AI \cite{facebook_map_with_ai} and Microsoft \cite{Microsoft_RoadDetections} have improved the completeness of global road network geometries, while parallel efforts from Microsoft\cite{microsoft_building_footprints} and Google\cite{google_open_buildings} have provided comprehensive inventories of building footprints. However, while these datasets have enhanced our understanding of the presence and location of global infrastructure, they do not capture the critical attributes that determine functionality and resilience. This semantic data gap—the absence of physical detail such as surface type—represents a major limitation in current infrastructure intelligence. The current de facto standard, OpenStreetMap (OSM), is a monumental achievement, yet its surface attributes are quite incomplete—covering only 30-40\% of the global network—and are often outdated\cite{haklay_quality_2010,su9060997,doi:10.1139/geomat-2021-0012}. A recent approach to assess road surface based on  street-level imagery \cite{RANDHAWA2025362} provided a large and relevant dataset but is fundamentally limited by the sparse coverage of open street-level imagery and by temporal inconsistencies - which renders the dataset unsuitable for a systematic global analysis. Consequently, our understanding of the planet's most essential infrastructure remains fragmented, static and misaligned with the pace of global change.

The convergence of high-cadence satellite remote sensing and advances in artificial intelligence has been described as a new paradigm to address challenges of data availability \cite{zhu_deep_2017}. The availability of global, high-resolution satellite imagery provides the necessary observational data, while deep learning models provide the capacity to extract and classify features at scale\cite{10.1145/3615900.3628772,cheng2022maskedattentionmasktransformeruniversal,DBLP:journals/corr/AlbertKG17,aleissaee2022transformersremotesensingsurvey,wang2022selfsupervisedlearningremotesensing,zhou_mapping_2024,rs15163985}. This combination overcomes the core limitations of both existing crowdsourced (OSM) and street-level data derived data products, by enabling a shift from a time agnostic data collection to active, dynamic monitoring. While prior efforts in combining remote sensing and AI have focused on mapping the geometry of infrastructure, our work addresses the next frontier: semantic enrichment—the extraction of physical and functional attributes that determine infrastructure performance and resilience. In this study, we leverage this technological convergence—albeit within the spatial constraints of 3–4 m PlanetScope imagery—to develop the first comprehensive, multi-temporal dataset of global road surface type, a first-order estimate of road width, and a derived Humanitarian Passability Score that integrates both attributes to assess functional accessibility across \textit{9.2 million kilometers} of the world’s critical \textit{arterial} network. 

In this study, we demonstrate that this multi-attribute dataset is more than an inventory, but a novel analytical lens through which to examine the geography of development, vulnerability, and equity across multiple spatial scales. It allows us to test the central hypothesis of this study: that the physical state and dynamics of infrastructure, when measured at high resolution, can serve as a powerful proxy for the complex and often invisible socioeconomic processes that shape our world. We investigate this hypothesis through a novel multi-scale analytical framework. At the \textit{planetary scale}, we build on foundational work linking satellite imagery to poverty\cite{doi:10.1126/science.aaf7894, doi:10.1126/science.abe8628}, introducing \textit{the rate of infrastructure change} as a dynamic metric that correlates with human development and offers near real-time insight into a nation's development trajectory—a perspective that is, by nature, complementary to traditional, slow-moving economic indicators\cite{yeh2020economic}. At the \textit{national scale}, we assess the functional consequences of road conditions by modeling network fragmentation and its implications for connectivity, trade, and economic vulnerability. At the \textit{local scale}, we use our data as a high-resolution lens to demonstrate how global infrastructure trends manifest spatially through case studies in Ghana and Pakistan. These applications show how our approach can expose the patterns of governance on urban equity and create actionable intelligence for climate resilience and humanitarian response.  

%Our work thus provides not only a foundational dataset but a scalable methodology for embedding infrastructure monitoring into global development research. It repositions road infrastructure from a static logistical asset to a dynamic, measurable indicator of economic resilience, social equity and local vulnerability —establishing it as a critical, actionable dimension of sustainable development.

This work establishes road surface data as a new class of high-resolution, physically interpretable proxies for human development—surpassing the spatial limits of traditional nighttime light–based approaches—and introduces a scalable framework for tracking the infrastructure foundations of prosperity, vulnerability, and climate resilience in a rapidly changing world.

%In this study, we first demonstrate that our methodology dramatically improves the completeness and temporal accuracy of global road surface data compared to existing sources. We then show that this dataset is more than an inventory; building on foundational work that links satellite imagery to poverty and sustainable development\cite{doi:10.1126/science.aaf7894, doi:10.1126/science.abe8628}, we establish it as a powerful tool for socioeconomic analysis. Specifically, and advancing beyond static assessments to capture temporal dynamics\cite{yeh2020economic}, we reveal that the \textit{rate of change} in road pavedness serves as a robust, high-resolution proxy for the Human Development Index (HDI), enabling a new capacity to monitor the pace of development in near real-time. Furthermore, we introduce a novel network-level analysis that reframes unpaved roads as a measure of economic fragility, quantifying the critical role they play in maintaining national connectivity. We reveal how our data can expose the spatial outcomes of national governance by mapping infrastructure patterns in Ghana, linking them to fragmented transport policies and intra-urban development inequities. In parallel, we use a case study from the 2025 Pakistan floods to showcase the data's utility in creating actionable intelligence for climate resilience and humanitarian response. Our work provides a foundational dataset and a new multi-scale framework for analyzing global infrastructure, from the dynamics of national development to the realities of local policy, vulnerability, and equity.

\section*{Results}

%\subsection*{Global Patterns of Road Pavedness and Infrastructure Dynamics}
\subsection*{A Multi-Temporal, High-Resolution Global Baseline of Road Infrastructure}

\begin{figure}
    \centering
    \includegraphics[width=1\linewidth]{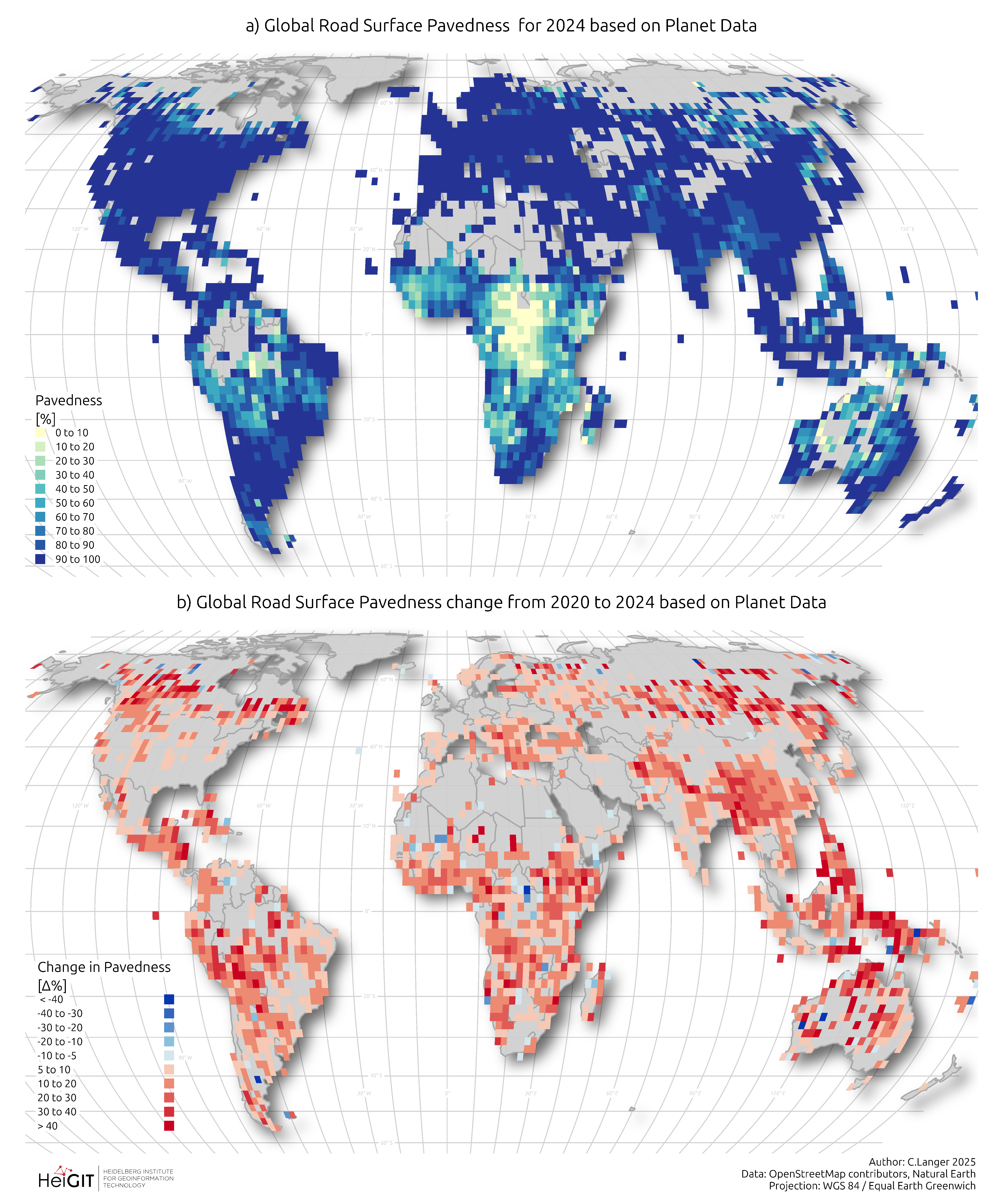}
    \caption{Global Road Infrastructure Status and Dynamics: Pavedness in 2024 and Change from 2020. (a) The upper panel displays the percentage of road length classified as paved within a global grid for primary arterial roads in 2024, based on deep learning analysis of Planet satellite imagery \cite{planet}. (b) The lower panel quantifies the absolute change in the share of paved roads between 2020 and 2024. Red cells indicate a net increase in pavedness, highlighting significant and rapid infrastructure investment, particularly concentrated in developing regions. Blue cells indicate an apparent decrease.}
    \label{fig:Pavedness_Urban_Rural_2024}
\end{figure}
Figure~\ref{fig:Pavedness_Urban_Rural_2024}(a) displays the state of road pavedness in 2024, revealing distinct continental-scale patterns. Large regions of the critical \textit{arterial} roads (see\hyperref[sec:def_arterial]{~\ref*{sec:def_arterial} definition})  (Figure ~\ref{fig:Pavedness_Urban_Rural_2024}) across North America, Europe, and East Asia exhibited high to complete paved ratios (>80\%), a finding consistent with our previous work showing these major road types are almost universally paved \cite{RANDHAWA2025362}. A key observation was the stark urban-rural dichotomy: even in highly developed regions, rural networks were less paved. This disparity was more pronounced in developing nations. Notably, a belt of low pavedness was evident across Central Africa and parts of South America. The disaggregation of this data for urban and rural road networks is presented in Supplementary Figure~\ref{S-fig:Pavedness_Grid_urban_rural}.

Moving beyond this static snapshot, we leveraged the temporal capabilities of our methodology to quantify infrastructure dynamics between 2020 and 2024. The resulting global change map (Figure ~\ref{fig:Pavedness_Urban_Rural_2024}(b)) reveals a predominantly positive shift in pavedness globally, confirming widespread investment in road improvement. However, this progress was uneven. The most significant increases (deep red hues) were concentrated in developing regions across South America, Africa, parts of South Asia, and Southeast Asia, visually pinpointing global hotspots of rapid road development. While the overarching trend was positive, localized areas of apparent decrease in pavedness (blue cells) were also present. These instances may represent model uncertainty due to variations in satellite imagery, but could also indicate actual road degradation or the construction of new unpaved roads, warranting further investigation. Together, these maps provide an unprecedented, high-resolution baseline and dynamic view of the world's arterial road network.

\begin{comment}
\begin{figure}
    \centering
    \includegraphics[width=1\linewidth]{pavedness_total_change_2020_2024}
    \caption{\textbf{Global hotspots of road infrastructure development from 2020 to 2024}. This figure quantifies the temporal change in road pavedness over a four-year period using our deep learning model on Planet satellite imagery. a, The map displays the absolute change in the share of paved roads between 2020 and 2024. Red cells indicate a net increase in pavedness, highlighting significant and rapid infrastructure investment, particularly concentrated in developing regions across Africa, South America, and South Asia. Blue cells indicate an apparent decrease. b, c, The static pavedness maps for 2020 (b) and 2024 (c) provide the temporal baseline and final state for the change analysis, visually confirming the overall increase in global road quality. \textit{For all panels, pavedness is calculated as the percentage of total road length (Paved / [Paved + Unpaved]) that is classified as paved.}}
    \label{fig:pavedness_total_change_2020_2024}
\end{figure}
\end{comment}

These regional and urban-rural disparities are quantified in Table ~\ref{tab:wb_regions_pavedness}. Developed regions such as Europe \& Central Asia and North America show near-complete paved networks (97.4\% and 96.9\%, respectively). In stark contrast, Sub-Saharan Africa averages only 63.1\% pavedness, with a significantly larger variance (±22.6\%) indicating high intra-regional heterogeneity. The data further reveals that the primary driver of this disparity is rural infrastructure; while urban road networks are largely paved across all regions (>93\%), rural pavedness in Sub-Saharan Africa (61.4\%) lags dramatically behind that of Europe \& Central Asia (97.2\%).

At the national level, these trends were even more pronounced (Supplementary Table ~\ref{S-tab:countries_pavedness_positions} ). Small, high-income nations such as Qatar, Barbados, and several Western European countries, exhibited virtually complete paved networks. Conversely, the countries with the lowest paved ratios were predominantly located in Sub-Saharan Africa, such as the Central African Republic (9.3\%), the Democratic Republic of Congo (19.0\%), and South Sudan (25.3\%). Notably, even among these nations, urban pavedness remained comparatively high, underscoring that the critical infrastructure deficit was overwhelmingly concentrated in rural areas. The low overall pavedness observed across the Central African belt may partly reflect challenging imaging conditions—such as dense forest canopy occlusion or road widths below the sensor’s effective resolution—which led to a higher proportion of roads being classified by our model as having an ‘unknown’ surface type.

\begin{figure}[htbp]
    \centering
    \includegraphics[width=\textwidth]{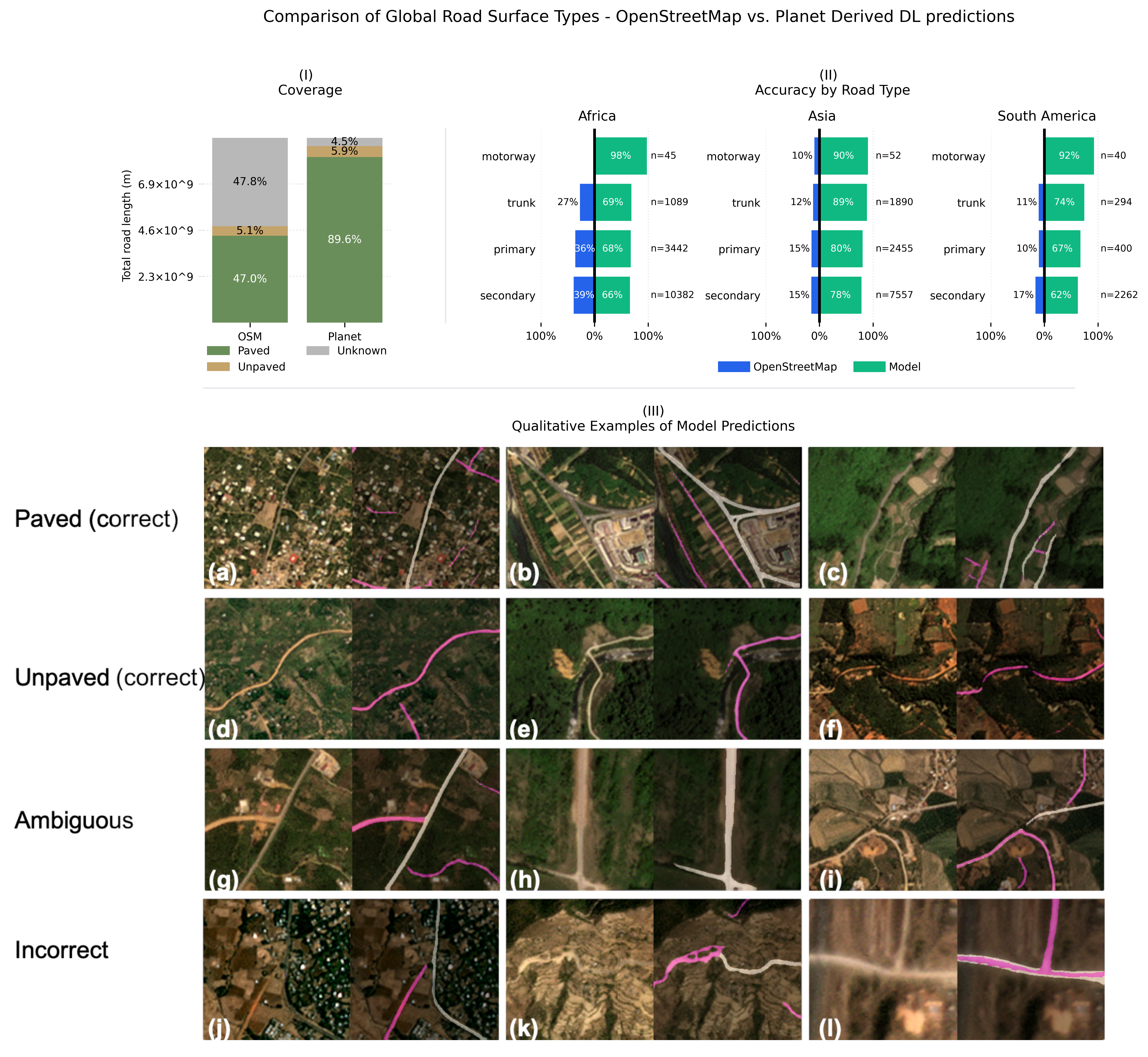}
    \caption{\textbf{Deep learning model accurately maps road development where OpenStreetMap lags}.
Top Panel (I) Comparison of coverage between OpenStreetMap (OSM) and our deep learning (DL)-derived Planet dataset \cite{planet}, showing the proportion of paved, unpaved, and unknown roads by total road length.(II) Accuracy by road type across three continents (Africa, Asia, South America), comparing OSM surface tags with predictions from our DL model against human-validated ground truth.
Bottom Panel (III) (i): Qualitative examples of model performance on Planet imagery \cite{planet}. For each example pair (a–l), the left image is the original satellite view  (Image © 2024 or 2020 Planet Labs PBC), and the right image is overlaid with the DL model's prediction (magenta for unpaved, white for paved). The first two rows show correctly classified paved (a–c) and unpaved (d–f) roads. The third row (g–i) highlights challenging cases, including an outdated OSM tag correctly updated by the model to 'paved' (g), a visually ambiguous compacted road (h), and a road with mixed surfaces (i). The fourth row (j–l) displays examples of false predictions by the model.
Right Panel (ii): Quantitative accuracy comparison. The charts compare the accuracy of existing OSM surface tags (blue) against our DL model's predictions (green) for major road types in Africa, Asia, and South America. The model consistently demonstrates substantially higher accuracy across all regions and road classes, underscoring its ability to provide a more current and reliable assessment of road infrastructure.
}
    \label{fig:osm_dl_accuracy}
\end{figure}

The model clearly outperformed OpenStreetMap (OSM) surface tags information by a wide margin (overall accuracy 89.2\% vs. 64.7\%) when validated against ground truth information (cf. Figure~\ref{fig:osm_dl_accuracy}\textit{ (Top Panel (II))}).
%A key validation of our approach is its demonstrated advancement over existing data sources in terms of both spatial coverage and predictive accuracy, as presented in Figure~\ref{fig:osm_dl_accuracy}\textit{ (Top Panel)}. Benchmarking against human-validated ground truth reveals that our deep learning model attains an overall accuracy of 89.2\%, outperforming OpenStreetMap (OSM) surface tags (64.7\%) by a wide margin.
This performance gap was consistent across all analyzed continents and road types (cf. Top Panel (II) Fig.~\ref{fig:osm_dl_accuracy}). Some qualitative examples (Bottom Panel (III) Fig.~\ref{fig:osm_dl_accuracy}) provide a clear explanation for this discrepancy: our model successfully captured recent infrastructure developments, such as newly paved roads that remained incorrectly tagged as 'unpaved' in the OSM database. The primary driver of this accuracy gap was presumably the rapid pace of development together with a low attention of OSM mappers to the road surface condition changes in regions where many real world features were not adequately mapped at all. A detailed breakdown of the model's performance, a discussion of validation challenges such as visually ambiguous road surfaces, and the full accuracy metrics are provided in the Methods section.

\begin{table}
\caption{\textbf{Pavedness by World Bank Region and Area Type.} Values (\%) are provided as the  mean and - in parenthesis - the standard deviation.}
\label{tab:wb_regions_pavedness}
\begin{tabular}{llllr}
\toprule
 & Total & Rural & Urban & N countries \\
Region &  &  &  &  \\
\midrule
Europe \& Central Asia & 97.4 (3.6) & 97.2 (3.7) & 99.9 (0.1) & 56 \\
North America & 96.9 (3.4) & 96.6 (3.7) & 99.9 (0.0) & 3 \\
Middle East, North Africa, Afghanistan \& Pakistan & 96.2 (7.5) & 95.8 (8.3) & 99.5 (0.6) & 22 \\
Latin America \& Caribbean & 90.8 (11.0) & 89.9 (11.9) & 99.6 (0.9) & 42 \\
East Asia \& Pacific & 90.7 (12.4) & 89.8 (13.1) & 99.0 (2.6) & 37 \\
South Asia & 90.5 (7.8) & 89.6 (7.8) & 98.2 (1.1) & 6 \\
Sub-Saharan Africa & 63.1 (22.6) & 61.4 (23.2) & 93.8 (7.1) & 48 \\
\bottomrule
\end{tabular}
\end{table}

\begin{figure}
    \centering
Figure    \includegraphics[width=1\linewidth]{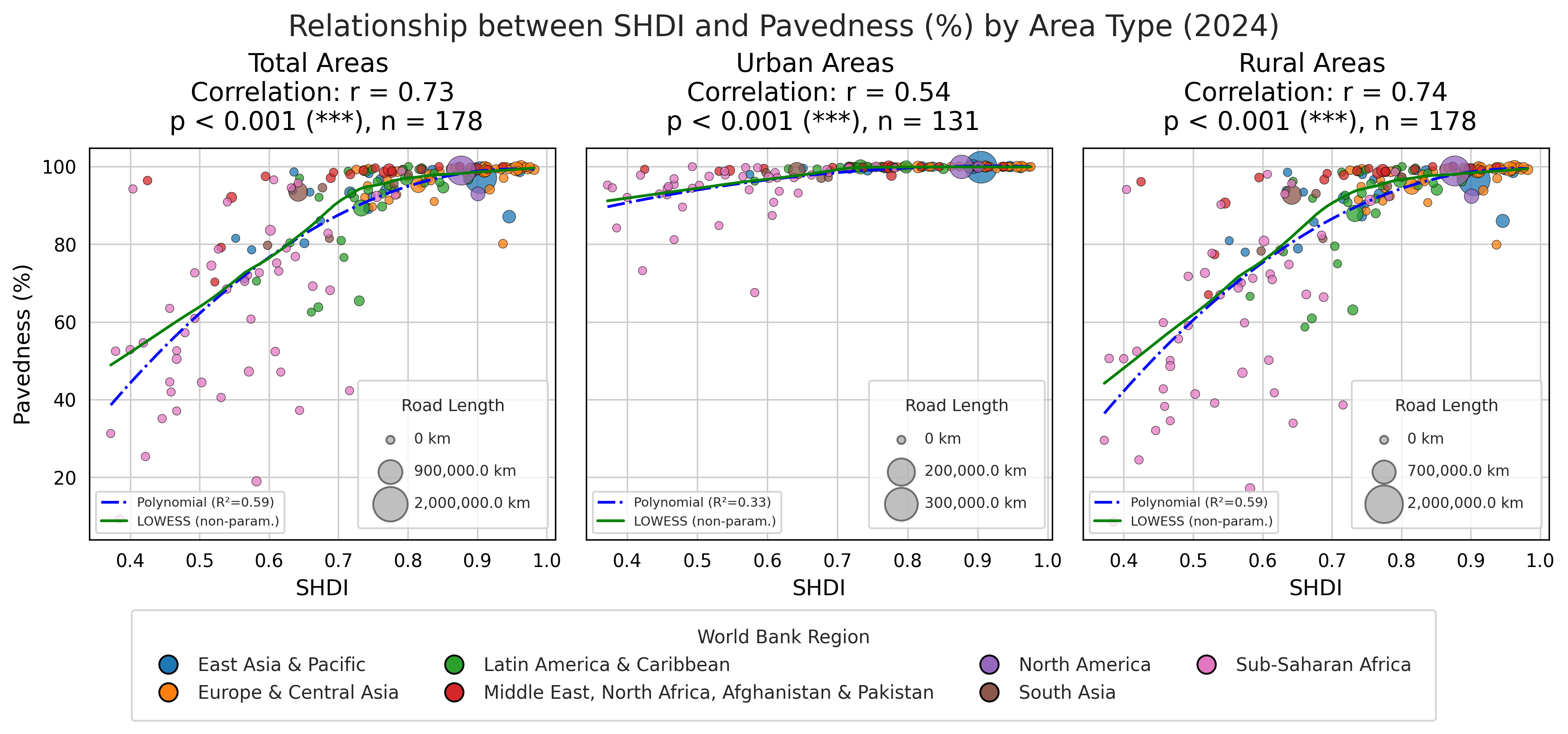}
    \caption{\textbf{Rural road infrastructure is a strong correlate of human development. }The figure shows the correlation between the percentage of paved roads in 2024 and the Subnational Human Development Index (SHDI). The relationship is disaggregated for the total road network (left), urban areas (center), and rural areas (right). Each point represents a country, colored by World Bank region and sized by the total road length within its borders.  The central panel reveals that urban road networks are near-saturated with high pavedness across most development levels. In contrast, the right panel shows a clear logistic-like relationship for rural roads, indicating that rural infrastructure quality is a more sensitive indicator of a country's development stage. Both polynomial (blue dashed line) and non-parametric LOWESS (green solid line) fits are shown.}
    \label{fig:pavedness-2024-hdi}
\end{figure}

\subsection*{Planetary-Scale Analysis: Road Dynamics as a High-Resolution Proxy for Human Development}

%To validate our dataset as a socioeconomic indicator, we correlated our pavedness metrics with the Subnational Human Development Index (SHDI). 

%To investigate the socioeconomic drivers of this disparity, we correlated our pavedness metrics with the Subnational Human Development Index (SHDI). 
Building on the established global infrastructure baseline, we used this dataset to evaluate our central hypothesis at the planetary scale—specifically, the relationship between road infrastructure dynamics and the Subnational Human Development Index (SHDI). The results demonstrate a consistent and statistically significant association: higher shares of paved roads in 2024 correspond to greater levels of human development, underscoring how the urban–rural infrastructure divide mirrors broader developmental gradients (Figure ~\ref{fig:pavedness-2024-hdi}).
The relationship is particularly pronounced for the total road network ($r = 0.73, p < 0.001$) and even stronger in rural regions ($r = 0.74, p < 0.001$)  while the correlation for urban areas is much lower ($r=0.54, p< 0.001$).

%The disaggregation exposes a critical insight: urban road networks are largely paved across most development levels, whereas rural road pavedness follows a distinct logistic-like growth curve, serving as a more sensitive barometer of a nation's development stage.

%While the static snapshot confirms the association between existing infrastructure and development, the temporal dimension of our dataset enables a more comprehensive and dynamic assessment. We therefore assessed the change in pavedness from 2020 to 2024, normalizing this change by each country's \textit{potential} for new paving (i.e., its share of unpaved roads in 2020). This metric, which represents \textit{the rate of infrastructure investment}, demonstrates a robust partial correlation with SHDI ($r_{\text{partial}} = 0.65$, $p < 0.001$), even after controlling for the baseline pavedness in 2020 (see Figure ~\ref{fig:potential-pavedness-hdi}). The strong linear relationship confirms that higher human development is associated with a faster rate of completing the national road network. This finding establishes that our remotely sensed road data can serve as a high-resolution, dynamic proxy for human development, capturing near real-time insights that are complementary to traditional, time-lagged official statistics. 

\begin{figure}
    \centering
    \includegraphics[width=1\linewidth]{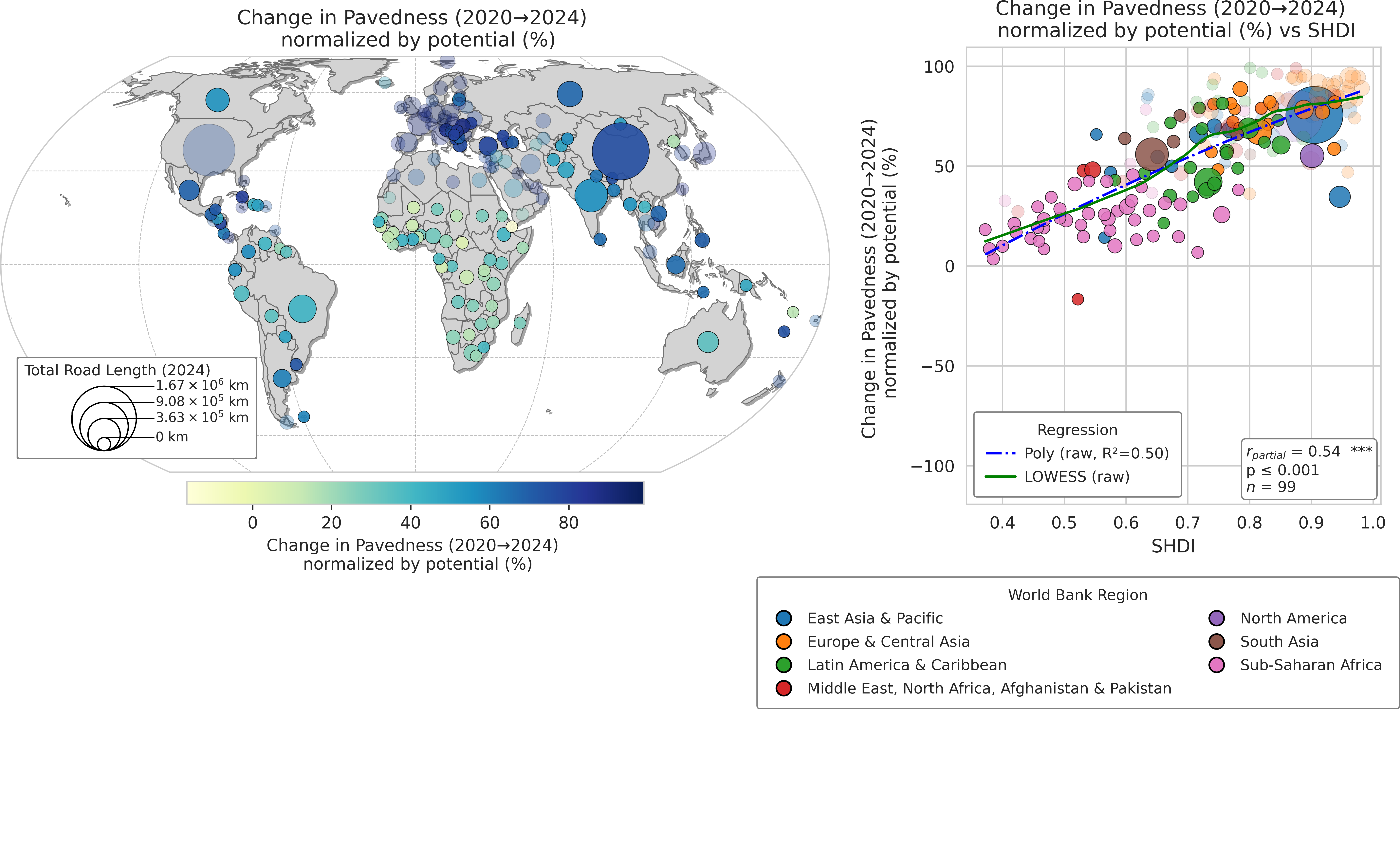}
    \caption{\textbf{The rate of road infrastructure improvement is strongly linked to human development.} This figure analyzes the dynamic relationship between the change in road pavedness from 2020 to 2024 and the Subnational Human Development Index (SHDI). To represent a standardized rate of infrastructure investment, the change in pavedness is normalized by the \textit{potential }for new paving (i.e., the share of unpaved roads in 2020). The left panel maps this normalized change rate globally, with each country's circle sized according to its total road network length. Countries with a baseline pavedness above 90 \% and less than 100 km of unpaved roads are shown with slight transparency to indicate their limited potential for additional paving. The right panel plots this normalized change against SHDI, including only countries meeting the relevance criteria. Despite this filtering, the dynamic metric reveals a strong and statistically significant partial correlation with SHDI ($r_{\text{partial}} = 0.65$, $p < 0.001$) after controlling for the baseline pavedness in 2020. This indicates that countries with higher human development are completing their paved road networks at a faster rate, independent of their starting point, thereby establishing this dynamic metric as a robust proxy for ongoing development. }
    \label{fig:potential-pavedness-hdi}
\end{figure}

While the static snapshot confirms the association between existing infrastructure and development, the temporal dimension of our dataset enables a more comprehensive and dynamic assessment. We therefore evaluated the change in pavedness from 2020 to 2024, normalizing this change by each country’s potential for new paving (i.e., its share of unpaved roads in 2020). This normalization emphasizes relative rates of improvement rather than absolute increases. Consequently, countries with already near-complete paved networks—such as most EU member states—may exhibit high normalized values even when the absolute change in paved length is small, reflecting maintenance cycles, localized construction, or model uncertainty rather than substantive expansion. To mitigate these effects, we filtered out countries with pavedness above 90 \% and less than 100 km of unpaved roads, which are shown with slight transparency in the global map (Figure ~\ref{fig:potential-pavedness-hdi}). Despite these exclusions, the normalized change in pavedness remains strongly correlated with SHDI ($r_{\text{partial}} = 0.65$, $p < 0.001$) after controlling for baseline pavedness. This finding establishes that our remotely sensed road data can serve as a high-resolution, dynamic proxy for human development, capturing near real-time insights that are complementary to traditional, time-lagged official statistics. 

%Unlike official statistics, which are often time-lagged and have coarse spatial resolution, our vector-based road data can provide granular, near real-time insights into development dynamics at subnational scales, offering an invaluable tool for organizations like the World Bank to monitor investment impacts and identify localized development gaps.

%- Change in Pavedness (2024 - 2020) divided through the potential (Percentage of that road network that can be paved). ->  This normalizes the change. For the Correlation to the HDI, we needed to account for the newly introduced dependency of the variable to the pavedness in 2020 -> Calculated the Partial correlation controlling for the pavedness in 2020 

% TODO rework figure to show bubble plots instead. Fix calculations for FR india china russia and us 

%\subsection*{National-Scale Analysis: Unpaved Roads as Critical Links for National Connectivity and Economic Resilience}

\subsection*{National-Scale Analysis: Network Integrity and Economic Fragility}

\begin{figure}
    \centering
    \includegraphics[width=1\linewidth]{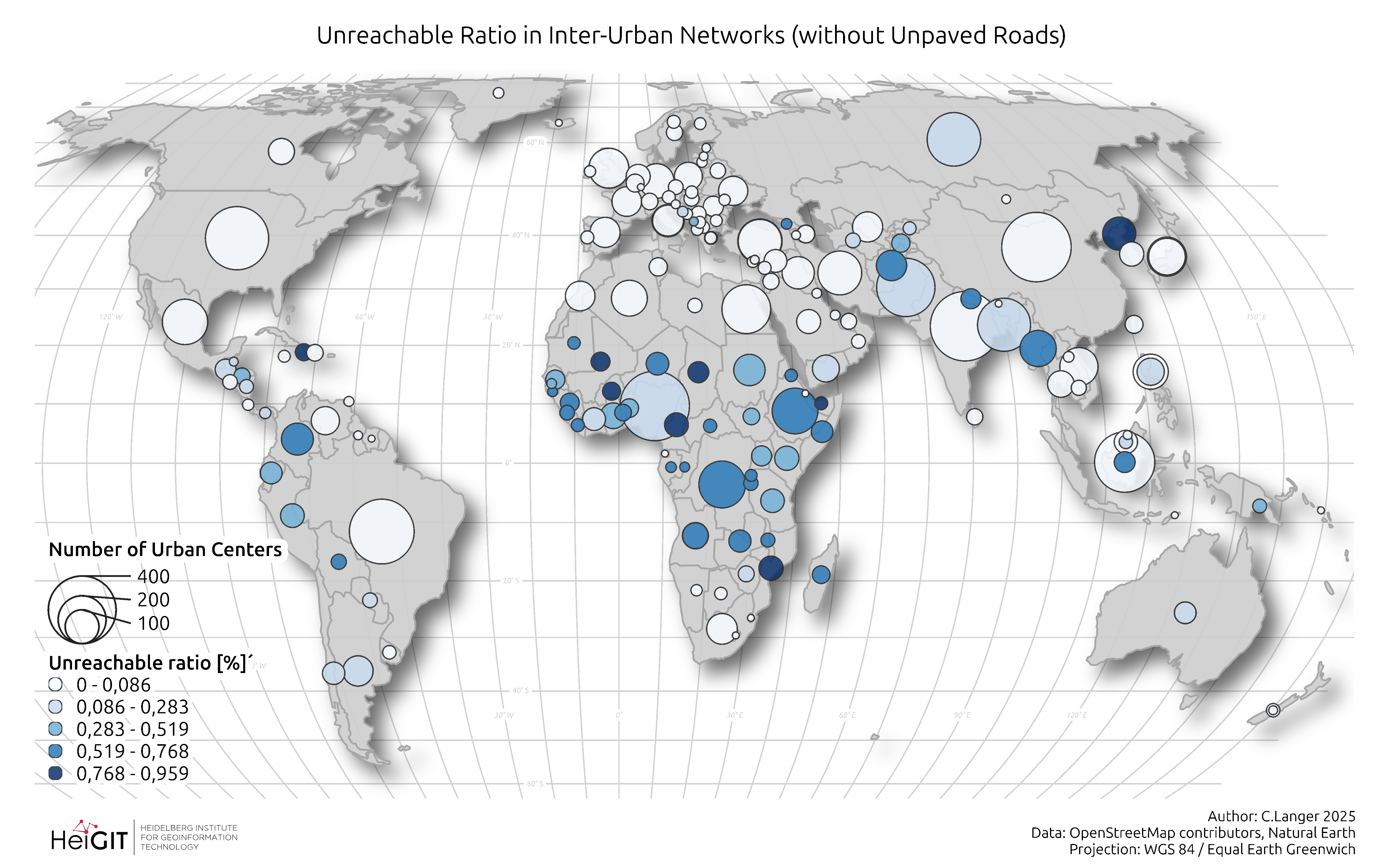}
    \caption{\textbf{A functional analysis of road networks reveals global hotspots of infrastructure fragility}. The figure quantifies the functional integrity of each country's road network by mapping the \textit{unreachable ratio} or share of inter-urban connections that become impossible when unpaved roads are removed from the transportation graph. Each country is represented by a circle, where the size is proportional to its number of urban centers and the color indicates the magnitude of connectivity loss. Countries with a low share of severed connections (0–0.1, white circles) exhibit a resilient, paved network that connects all major economic hubs. In contrast, high shares (dark blue) reveal a fragmented paved network where unpaved roads form critical, non-redundant links. In doing so, this metric transcends conventional pavedness indicators to offer a direct, structural measure of economic exposure—pinpointing nations where infrastructure gaps critically undermine trade resilience and national connectivity.}
    \label{fig:connectivity_urban_centers}
\end{figure}

After establishing a global relationship between road dynamics and development, we extended our analysis to the national scale to quantify the functional impact of infrastructure quality on economic connectivity. Leveraging our dataset, we performed a systems-level assessment of national road network integrity, advancing beyond correlation-based analyses to evaluate how incomplete networks constrain economic performance. We conducted a graph-based connectivity analysis by simulating the removal of all unpaved roads and calculating an \textit{unreachable ratio}—the share of trips between major urban centers that become impossible.
The results, visualized in Figure ~\ref{fig:connectivity_urban_centers}, reveal the critical and often overlooked role that unpaved roads play in maintaining national connectivity. The map exposes a stark global divide. In developed nations across North America, Europe, and East Asia, the unreachable ratio is near zero. This signifies that their paved road networks are mature and resilient, providing multiple, redundant paths between economic hubs; here, unpaved roads primarily serve local access. Conversely, in many nations across Sub-Saharan Africa, and parts of South America and Southeast Asia, the ratio is high (light to dark blue circles). This indicates a high dependency on unpaved roads for strategic, inter-city transport. In these countries, the unpaved network is not merely for peripheral use but forms the essential connective tissue that holds the national economy together. This analysis therefore provides a direct, quantitative measure of a nation's dependency on its unpaved road network, revealing a geography of infrastructural fragility.

\subsection*{From Global to Local: High-Resolution Insights for Policy, Climate Vulnerability and Development Equity}

Having established the planetary and national-scale dynamics of infrastructure, we now demonstrate the potential of this dataset to bridge from these broad patterns to the local realities that shape human lives. This final stage of our multi-scale analysis uses high-resolution case studies to reveal how our data can serve as a lens to investigate the on-the-ground drivers and consequences of infrastructure inequality, from the outcomes of fragmented governance to the tangible risks of climate shocks.

\subsubsection*{Case Study Ghana: Multi-Scale Analysis of Infrastructure Equity and Development}

\begin{figure}
    \centering
    \includegraphics[width=1\linewidth]{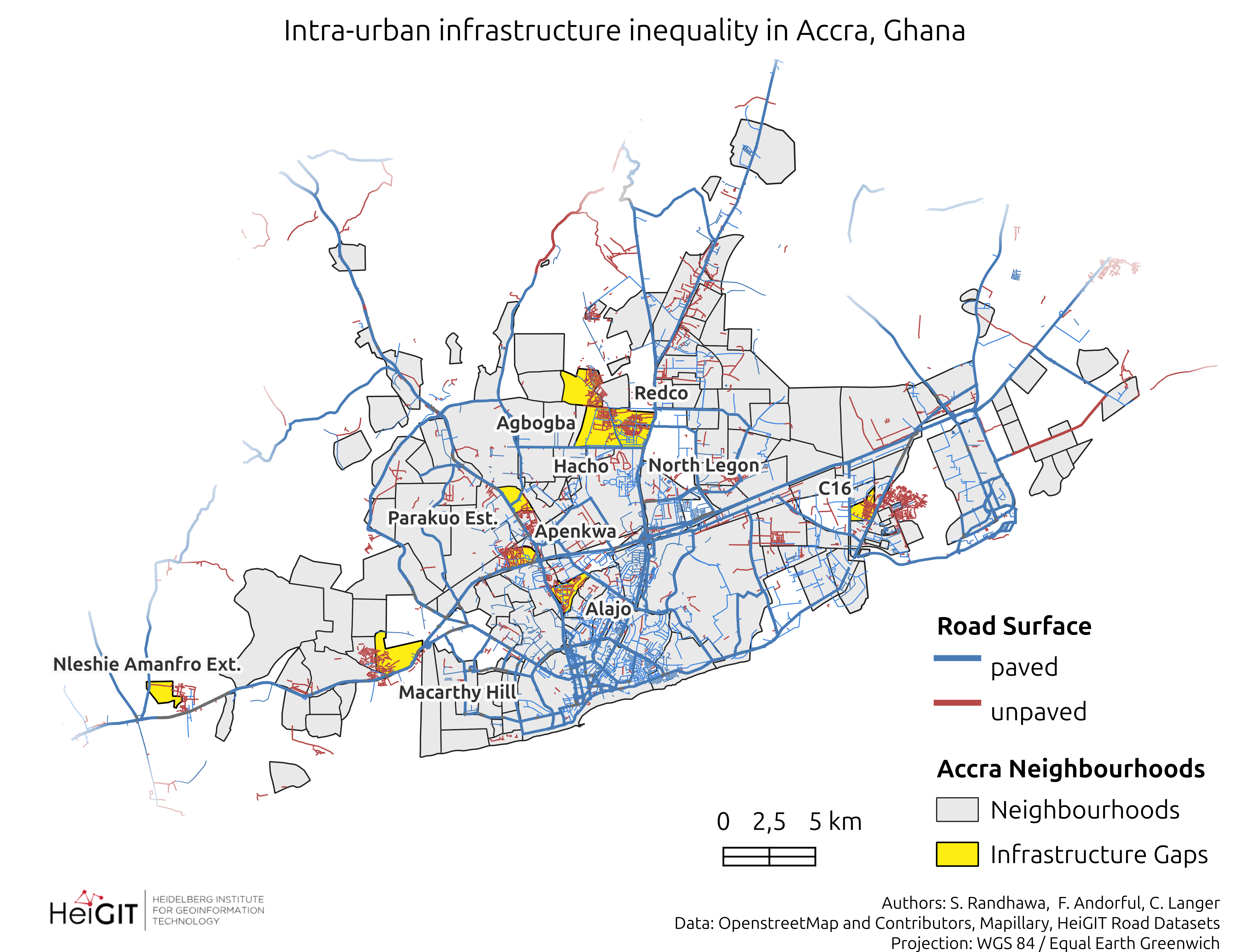}
    \caption{\textbf{Intra-urban infrastructure inequality in Accra, Ghana.} This map visualizes road surface conditions across Accra's neighborhoods, combining our DL-model-derived data with previous Mapillary-based classifications to create a comprehensive view. The analysis highlights 'Infrastructure Gaps' (yellow areas), such as Redco and Macarthy Hill, where a high density of unpaved roads indicates lagging development compared to more central, well-serviced neighborhoods. This granular, neighborhood-level view is critical for equitable urban planning and resource allocation.}
    \label{fig:accra_inequality}
\end{figure}

\begin{figure}[htbp]
    \centering
    \includegraphics[width=\textwidth]{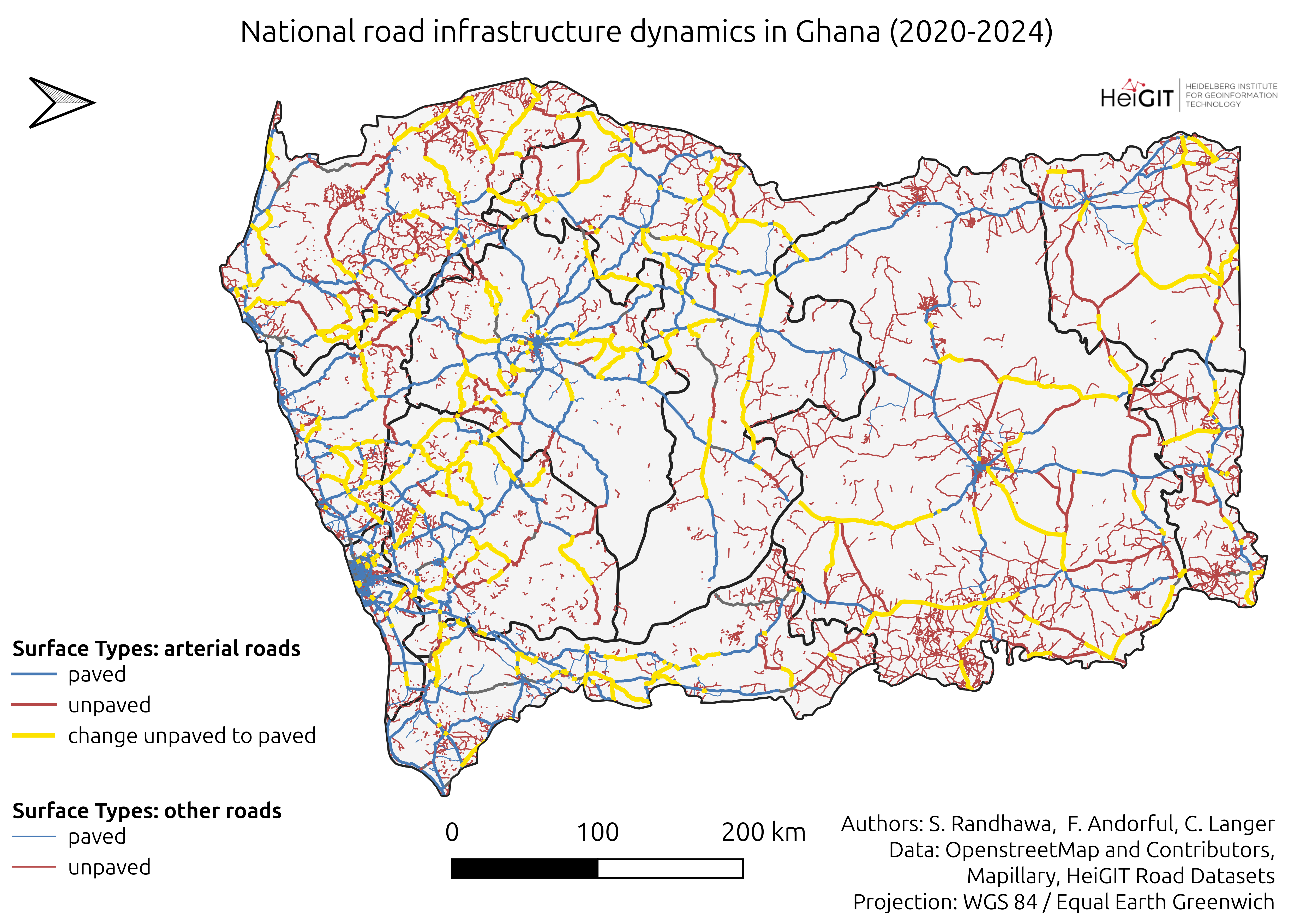}
    \caption{\textbf{National road infrastructure dynamics in Ghana (2020-2024).} The figure displays the state of Ghana's arterial road network. \textbf{a}, The network state in 2024, showing existing paved (blue) and unpaved (red) roads. \textbf{b}, The network development between 2020 and 2024, highlighting newly paved roads (yellow) which represent infrastructure investments over the four-year period. The map reveals distinct regional patterns, with a higher concentration of unpaved roads remaining in the Southwest and Northeast. The distribution of newly paved roads provides a transparent record of development priorities, serving as an evidence-based tool to assess the implementation of national transport policy.}
    \label{fig:ghana_development}
\end{figure}

To showcase the multi-scale utility of our dataset for subnational analysis, we conducted a case study of Ghana, examining infrastructure patterns at both the urban and national levels. At the urban scale, we focused on the capital, Accra, combining our Planet-derived data with previous street-view classifications\cite{RANDHAWA2025362} to create a comprehensive infrastructure map (Figure~\ref{fig:accra_inequality}). This analysis reveals significant disparities in road quality between neighborhoods. The map allows for the clear identification of "Infrastructure Gaps"—areas such as Redco and Macarthy Hill where a high density of unpaved roads indicates lagging development and potential inequities in resource allocation compared to more central, well-serviced neighborhoods. Other interesting observation is neighborhoods like Sakumono (C16) where estate developers operate shows estates like Sakumono estate fully paved while its 360-degree surrounding neighborhoods are unpaved. This granular, neighborhood-level view is critical for data-driven urban planning and for addressing intra-city inequality.
At the national scale, the dataset provides a data-driven tool for evaluating the implementation of transport policy (Figure~\ref{fig:ghana_development}). The map of Ghana's arterial roads reveals clear patterns, with a higher concentration of paved roads on the trunk roads, with the Southwest and Northeast more unpaved. At the same time, intercity roads are generally unpaved except the capital (Accra) and the second largest city (Kumasi). Crucially, the temporal data from 2020 to 2024 highlights specific road segments that have been recently paved, offering a transparent record of where infrastructure investment has been focused. 

This pattern potentially is grounded in the road governance architecture in Ghana: The Ministry of Roads and Transport formulates policies, coordinates sector performance, and oversees road infrastructure development \cite{MRH}. However, planning, construction, and maintenance are delegated to three largely autonomous agencies with distinct mandates: (1) Ghana Highways Authority (GHA), responsible for trunk roads \cite{GHA}; (2) Department of Urban Roads (DUR), managing urban networks \cite{DUR}; and (3) Department of Feeder Roads (DFR), tasked with feeder roads linking major settlements to trunk routes \cite{DFR}. This administrative separation might explain the visible road hierarchy. GHA, overseeing economically vital and high-traffic routes, naturally secures larger investments and political attention. Its projects are high-cost and high-visibility, whereas DUR and DFR projects, though critical to local communities, are smaller, more numerous, and more prone to budget shortfalls. The uneven quality of paved versus unpaved roads might thus be a spatial reflection of this fragmented governance. Recognizing these inefficiencies, the government submitted the National Roads Authority (NRA) Bill, 2023, acknowledging that the “lack of coordination and separate mandates” among agencies is a major impediment to progress and proposing their consolidation \cite{BILL}.

At the urban scale, the dynamics of unpaved community roads in the Accra - regardless the higher concentration of paved roads - presumably depicts the impact of its urban growth. Between 1985 and 2020, the city’s urban footprint expanded by nearly 400\%, far outpacing transport infrastructure and exposing weaknesses in coordination \cite{cross_road}.
The Sakumono case (C16) shows an interesting factor where private developers pave roads as a market-driven strategy to increase property value, operating under a different set of financial incentives than public agencies.

Recent government initiatives, such as the "Year of Roads" programs between 2020 and 2024,\cite{YOR} presumably have spurred construction activity, particularly on strategically important and economically critical trunk roads and interchanges. This presumably contributed to the changes that we see in 2024.
This kind of dynamic, data-oriented overview can serve as a "indicator" providing objective assessment of how effectively initiatives align with national strategic priorities and whether it achieves equitable development across historically under-served regions.  Allowing for an objective assessment of whether development is aligned with national strategic priorities and reaching historically neglected areas.

\subsubsection*{Case Study Pakistan: Infrastructure Vulnerability and Climate Shock Resilience}

Through another detailed case study in a flood-affected region of Pakistan, we demonstrate the dataset’s utility in generating actionable intelligence for disaster response and climate adaptation, leveraging flood extent data from the Copernicus Emergency Management Service (EMS). This analysis demonstrates how addressing critical infrastructure shortfalls can translate into refined, operational intelligence that informs resilience strategies.

\begin{figure}[htbp]
    \centering
    \includegraphics[width=\textwidth]{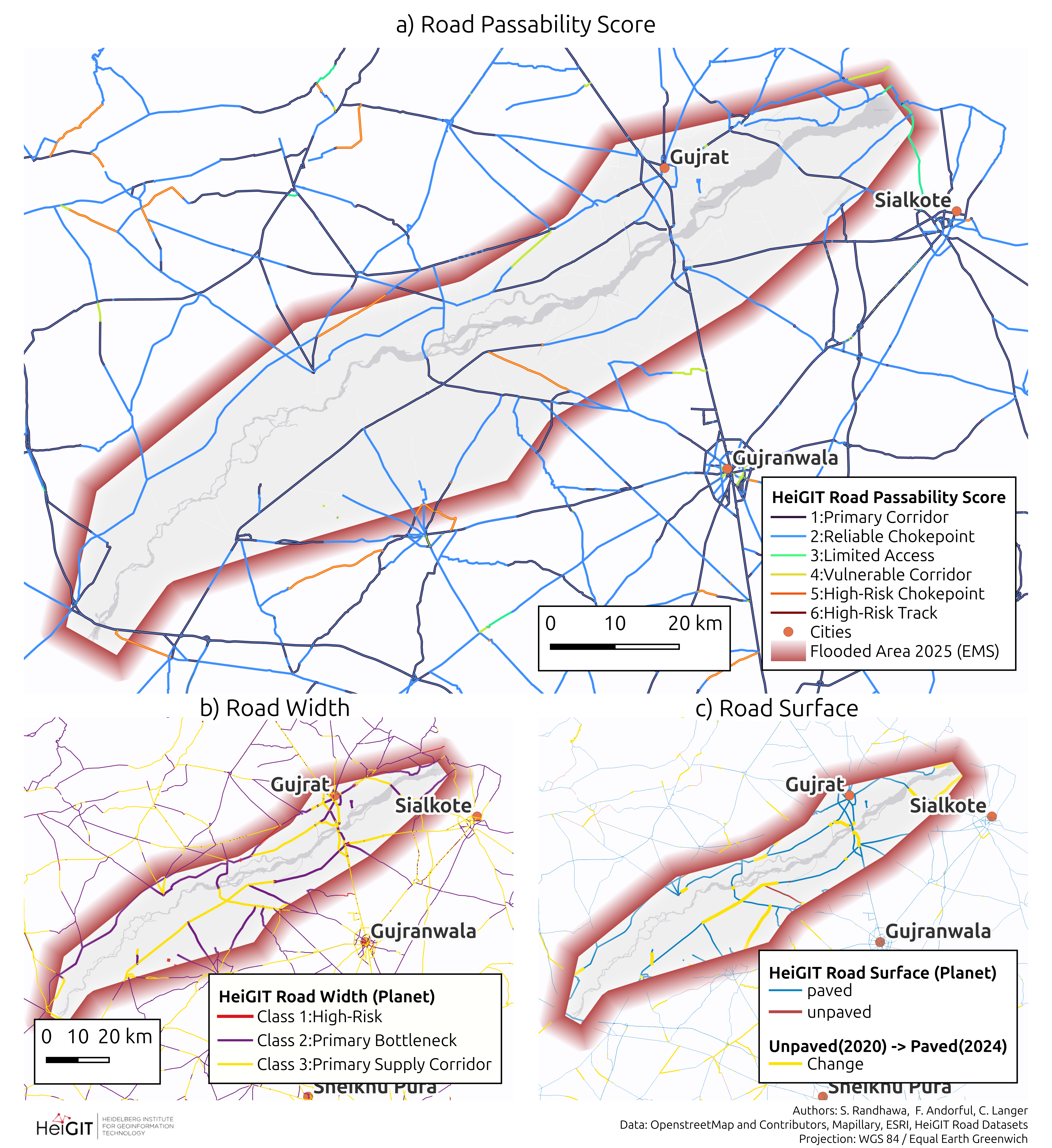}
    \caption{\textbf{High-resolution infrastructure analysis in a flood-prone region of Punjab, Pakistan reveals climate vulnerability and informs humanitarian response.} The figure demonstrates an end-to-end workflow from data creation to actionable intelligence within a region impacted by flooding (indicated by the dashed boundary from Copernicus EMS). \textbf{a Road Passability Score:} Synthesizes the derived road attributes into a Humanitarian Passability Score, providing an immediate, color-coded assessment of network resilience and risk for disaster planning. It identifies primary corridors, vulnerable areas, and high-risk choke points within the flooded region. \textbf{b Road Width Classification:} Displays the derived road width classified into logical categories: Class 1 (High-Risk), Class 2 (Primary Bottleneck), and Class 3 (Primary Supply Corridor), which are crucial for understanding network capacity and potential choke points. \textbf{c Road Surface Status and Temporal Change:} Displays the road surface conditions within the flood-affected region based on our deep learning analysis of Planet imagery. Light blue lines represent roads classified as paved in 2024, red lines indicate unpaved roads in 2024, and yellow lines highlight newly paved roads that transitioned from unpaved in 2020 to paved in 2024, illustrating recent infrastructure development and dynamics in the region.}
    \label{fig:pakistan_case_study}
\end{figure}

First, we highlight the significant infrastructure data gap in existing public datasets for this region. As visualized in Supplementary Figure~\ref{S-fig:Climatshock_appendix}, the road surface information derived from our DL-model provides near-complete coverage, in stark contrast to the sparse data available from the combined OSM and Mapillary datasets, a pervasive challenge in South Asia. Beyond filling this gap, our methodology captures the temporal dynamics of development. Figure~\ref{fig:pakistan_case_study}c displays the road network's state in 2024, revealing both the existing paved and unpaved roads, as well as the significant recent investments shown by the newly paved roads from 2020 to 2024.

Furthermore, we enriched this analysis by deriving road width from the segmentation masks. Despite being a first-level estimate influenced by imagery resolution, (Figure~\ref{fig:pakistan_case_study}b) vividly classifies the road network's width, revealing that many segments are primary bottlenecks and primary supply corridors, interspersed with critical high-risk narrow roads —a significant constraint for logistics. Ultimately, we synthesize the surface and width data into a final "Humanitarian Passability Score" (Figure~\ref{fig:pakistan_case_study}a). This score provides an immediate, actionable assessment of the network's resilience, identifying primary supply corridors versus high-risk chokepoints and tracks. This type of granular, multi-attribute analysis is pivotal for reducing lead times in emergency planning and for strategizing long-term climate adaptation investments to strengthen the most vulnerable links in the infrastructure network.

\section*{Discussion}

This study presents the first comprehensive, dynamic, and multi-attribute global dataset of road infrastructure derived from (Planetscope (3-4m)) satellite imagery. By moving beyond outdated and incomplete inventories, our approach offers a novel lens for examining the intersection of infrastructure investment, human development, and economic resilience across spatial and temporal scales. The analysis spans from global patterns to context-specific, policy-relevant applications, demonstrating the dataset’s utility for both research and decision-making. The robust empirical relationships we identify between road quality and development outcomes align with established findings in development economics, while the high spatial resolution of our dataset enables this relationship to be investigated at a scale and granularity not previously possible\cite{Gorgulu2023Infrastructure}.

Previous efforts to map development dynamics have largely relied on nighttime light intensity as a proxy for economic activity \cite{doi:10.1126/science.aaf7894, yeh2020economic}, offering valuable—but spatially coarse—insight into development. In contrast, our dataset introduces road surface type as a high-resolution physical proxy that directly captures infrastructure investment and accessibility, thereby enabling more granular assessments of economic development, vulnerability, and climate resilience. This establishes road surface type as a critical and actionable dimension of sustainable development.

This work substantially augments the OpenStreetMap (OSM) database and marks a significant advancement over prior global infrastructure datasets, including our earlier efforts based on street-level Mapillary imagery\cite{RANDHAWA2025362}. Although that approach added valuable granularity, its limited spatial coverage led to only a marginal increase in global road surface information—from 33\% to 36\%. In contrast, the satellite-based methodology introduced here achieves \textit{near-complete coverage} for the 9.2 million km of critical arterial roads analyzed, effectively addressing a major data shortfall in which nearly half (47.8\%) of this network previously lacked any surface type classification (Figure~\ref{fig:osm_dl_accuracy}, Top Panel (I)).

By incorporating multiple time points, this dataset provides a dynamic, multi-temporal perspective on road infrastructure, capturing critical patterns of change. At the planetary scale, our findings establish a new paradigm for monitoring development. By quantifying the \textit{rate of change} in pavedness between 2020 and 2024, we demonstrate that this metric of infrastructure investment is a powerful proxy for a country’s active development process, correlating strongly with the Human Development Index. When viewed alongside traditional indicators such as nighttime lights, pavedness offers complementary insights into socioeconomic dynamics, providing a direct, infrastructure-based signal of development progress and resilience.This transforms our dataset from a descriptive map into a dynamic monitoring tool, offering a novel capability to track the pace of national development and assess the impact of infrastructure projects in near real-time. Specifically, our vector-based data can provide granular insights at subnational scales, providing international bodies like the World Bank and national governments, a powerful data-driven tool for evaluating progress towards goals, such as those outlined in the UN Sustainable Development Goals, bypassing the significant time lags of official reports.

At the national scale, our functional network analysis reconceptualizes unpaved roads—not merely as indicators of infrastructure deficit, but as markers of economic vulnerability. The \textit{unreachable ratio} starkly illustrates that for many developing nations, the unpaved network is a critical, fragile backbone for national connectivity. A high ratio implies that supply chains are fragile and economic integration is hampered. This metric provides a tangible, systems-level benchmark for policymakers, suggesting that strategic investment should prioritize paving the critical links required to unify the national economy.

The strength of this multi-scale framework is most evident at the local scale, where it reveals the underlying drivers and real-world consequences of these broader infrastructure patterns. Our case studies illustrate this in distinct contexts: in Ghana, the data uncovers how fragmented governance structures manifest as deep-seated inequities in urban road quality, while in Pakistan, our \textit{Humanitarian Passability Matrix} provides actionable intelligence for assessing climate vulnerability and strengthening humanitarian response logistics. This ability to bridge from global dynamics to local realities is a key contribution of our work.

A central finding of this study is that, when analyzed at high resolution, the rural road network serves as a sensitive indicator of a nation’s development stage. Our analysis quantifies the pronounced urban-rural divide and its strong correlation to socioeconomic status. This empirical, surface-aware approach offers a significant advancement over the widely used Rural Access Index (RAI), the standard metric for assessing rural connectivity\cite{Roberts_etal_2006}. Existing global RAI models rely heavily on OSM road information, often operating under simplifying assumptions about surface condition, thereby overlooking a critical factor in a network's true, all-weather functionality\cite{sun2023ruralaccessindexglobal,iimi2016ruralaccess,Sun_etal_2023}. By directly capturing surface conditions, we provide this critical missing dimension needed to interpret how accessibility and vulnerability co-evolve in space and time.

It is crucial, however, to interpret these findings not as a universal call for paving all roads, but as a framework for understanding strategic infrastructure trade-offs. While paved roads are strongly linked to higher development indices, unpaved roads retain a vital role in many regions for their cost-effectiveness and lower environmental impact, provided they are well-maintained\cite{african2014tracking, Yang2024, ESA2021}. Our dataset should therefore be seen as a high-resolution tool to help policymakers identify where strategic paving offers the greatest socioeconomic benefit, while also recognizing where well-maintained unpaved infrastructure remains the most sustainable solution.

Methodologically, this study underscores a critical insight for the big data era: while crowdsourced platforms like OSM are foundational, their attribute data often fails to keep pace with real-world infrastructure change\cite{herfort_spatio-temporal_2023}. Our human-validated assessment revealed that OSM surface tags are frequently outdated, achieving only 26\% global average accuracy on unpaved roads (Figure~\ref{fig:osm_dl_accuracy}, Top Panel (II)). Rather than replacing community mapping, our AI-driven approach serves as a powerful complement—offering a scalable mechanism to systematically enrich and update these globally important datasets.

Nevertheless, we acknowledge several limitations of this study. Model performance may be affected by challenging imaging conditions and regional variations in road typology, reflecting a strategic training focus on the Global South. Furthermore, our analysis is constrained by the 3-4m spatial resolution of the PlanetScope imagery, which affects the precision of our derived road width attribute and prevents the resolution of finer details like individual lanes (see Supplementary Figures \ref{S-fig:unknownroads_share} and \ref{S-fig:width_extraction_examples}). While we have established strong correlations, we do not claim direct causation between road paving and HDI. These limitations define clear avenues for future research, including developing regionally-specialized models to leveraging higher-resolution, multi-modal data to create a truly comprehensive infrastructure inventory and the design of causal frameworks to more precisely evaluate the economic impacts of targeted infrastructure investments.

Furthermore, future studies and connectivity analyses should move beyond traditional binary reachability metrics by incorporating distance-weighted measures such as travel cost metrics. These more nuanced indicators will offer richer insights into the functional impacts of infrastructure development.

\section*{Methods}

The methodology for generating the global road surface dataset was structured in two main parts: a streamlined description of the global dataset creation pipeline and a high-level overview of the deep learning model development. Comprehensive details regarding model training, architectural choices, and performance analyses are provided in the Supplementary Information. In the model training phase, we fine-tuned a semantic segmentation model (Mask2Former)\cite{cheng2022maskedattentionmasktransformeruniversal} on a curated dataset of PlanetScope imagery labeled with road surface types. In the global dataset creation pipeline, we developed a scalable approach to extract, process, and classify millions of road segments worldwide using OpenStreetMap geometries and Planet satellite imagery (RGB). The full dataset, with all attributes detailed in Supplementary Table~\ref{S-tab:dataset_attributes}, has been made publicly available through the Humanitarian Data Exchange (HDX) to support further research and operational applications.

\subsection*{Global Dataset Creation Pipeline}

\begin{figure}[htbp]
    \centering
    \includegraphics[width=\textwidth]{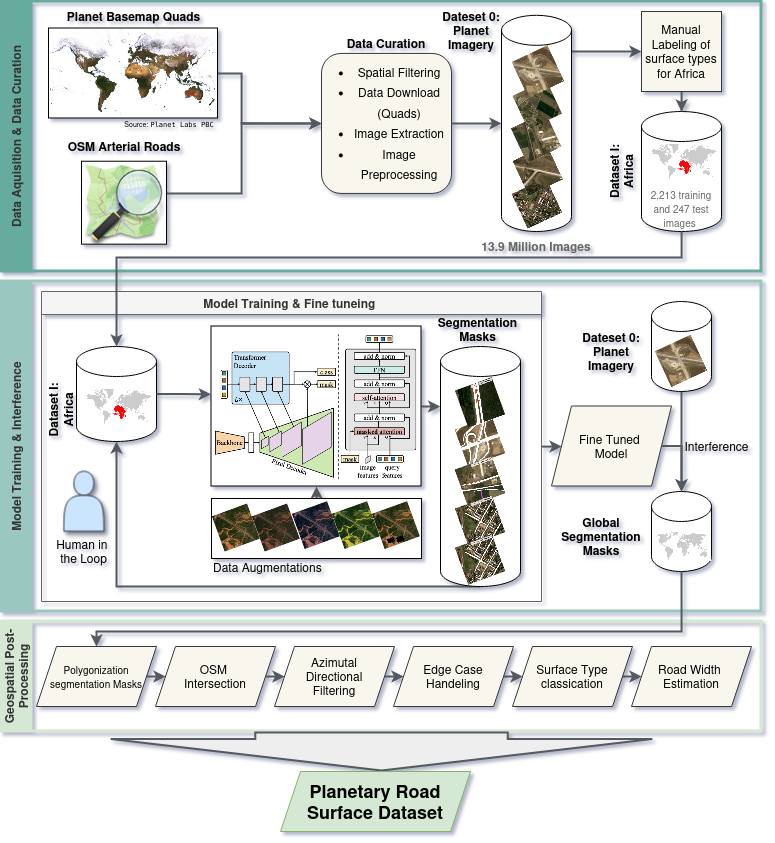}
    \caption{Deep Learning-Driven Pipeline for Global Road Surface and Width Classification.
An overview of the end-to-end workflow, from raw data to final product. The process initiates with Data Acquisition and Curation using Planet imagery and OSM road geometries, including extensive image preprocessing and manual labeling for African training data. This feeds into Model Training and Inference, where a deep learning segmentation model (Transformer Decoder) is fine-tuned and then applied globally to generate segmentation masks. The final Geospatial Post-Processing phase refines these masks through polygonization, integration with OSM, directional filtering, and advanced classification steps to yield the Planetary Road Surface Dataset, complete with surface type and road width information.
}
    \label{fig:road_surface_workflow}
\end{figure}

The global dataset creation pipeline involved large-scale sampling, imagery extraction, semantic segmentation, and geospatial postprocessing to classify road surfaces for millions of OpenStreetMap segments. A custom big data download workflow was implemented to overcome Planet API constraints, enabling efficient extraction of high-resolution PlanetScope imagery for approximately 13.9 million global road locations. Each road segment was processed through semantic segmentation followed by a series of spatial operations to assign a surface classification.

A schematic overview of this pipeline is shown in Figure \ref{fig:road_surface_workflow}, which summarizes the end-to-end flow from data acquisition and curation to the final classification of the road surface.

The first step in our analysis was to create and validate a comprehensive, dynamic baseline of global road infrastructure, overcoming the coverage and temporal limitations of existing datasets. To elucidate global road infrastructure patterns and dynamics, we aggregated our deep learning-based surface predictions into a standardized grid format (Figure~\ref{fig:Pavedness_Urban_Rural_2024}) and derived regional statistics (Table~\ref{tab:wb_regions_pavedness}, Supplementary Table~\ref{S-tab:countries_pavedness_positions}). 

Details on global geospatial sampling, PlanetScope imagery acquisition, semantic segmentation inference, and the geospatial post-processing pipeline, including its sub-steps and rationale, are provided in the Supplementary Methods.

\subsection*{Deep Learning Model Overview}
For the core task of classifying road surface conditions, we utilized a fine-tuned Mask2Former model for semantic segmentation. This pixel-wise classification approach was selected for its ability to accurately capture the irregular shapes, widths, and extents of road networks. The model was trained on a high-quality, manually labeled dataset derived from diverse regions, and further optimized using strategies such as targeted data augmentation and hard pixel mining to ensure robust global generalization.
Further comprehensive details on the deep learning model development, including the specific architecture, training protocols, data labeling process, benchmarking against other models, and strategies employed to address the generalization gap, are provided in the Supplementary Methods.

\subsection*{Quantifying Connectivity Loss Due to Unpaved Roads}

To quantify the functional role of unpaved roads and their impact on national connectivity, we developed a methodology to assess the share of inter-urban connections that become impossible when unpaved segments are removed from the transportation graph.

Initially, urban center polygons for each country were extracted and simplified to convex hulls, and corresponding OpenStreetMap (OSM) road networks were pre-cleaned and extracted per region using the osmium-tool to retain only arterial classes (e.g., motorways, primary, secondary roads) and subsequently analyzed with the OSMnx package \cite{OSMNX}.
For each urban center, we identified entry points by computing intersections between the center's boundary and road edges. To manage complexity, these candidate intersection points were reduced to a maximum of three representative "entry points" per urban center using k-means clustering \cite{hartigan1979algorithm}, excluding those belonging to small, isolated subgraphs. These selected entry points were then snapped to the nearest nodes in the OSM graph.
A baseline measure of inter-urban connectivity was then established. For each urban center, multi-source Dijkstra's algorithm \cite{dijkstra_1959} was run from its identified entry nodes across the full road network, using road length as the travel cost. This generated a symmetric urban-to-urban distance matrix, recording the minimum travel distance between entry nodes for every pair of centers. Pairs that were unreachable were marked accordingly.
Next, all road segments previously identified as "unpaved" in our predicted surface dataset were programmatically removed from the network. The connectivity analysis was then repeated on this pruned network. Any pair of urban centers that was connected in the baseline (full) network but became unreachable after the removal of unpaved roads was classified as a newly non-connectable pair. The final metric, representing the proportion of national connections dependent on unpaved roads, was calculated as the share of these newly non-connectable pairs relative to all originally reachable pairs. This metric thereby quantifies the reliance of each country's transportation system on unpaved infrastructure for maintaining inter-urban connectivity.

\subsection*{Analyzing Road Infrastructure Development and Human Development Linkages}
To investigate the dynamic relationship between road infrastructure development and human development, we analyzed country-level pavedness metrics in relation to the Subnational Human Development Index (SHDI) and World Bank regional classifications.
Pavedness was defined as the proportion of a country's total road network surfaced with durable materials (e.g., asphalt or concrete) relative to the sum of paved and unpaved road lengths. This metric was derived for total, urban, and rural areas from our aggregated road-length statistics for each country.
The relationship between SHDI and pavedness was explored through bubble scatterplots, where the area of each bubble proportionally represented the country's total road length. To capture various functional forms of this relationship, we applied  polynomial, and locally weighted scatterplot smoothing (LOWESS) regressions. The strength of the linear association was quantified using the Pearson correlation coefficient ($r$), which was preferred over Spearman's rank correlation due to the continuous nature of both SHDI and pavedness, and our focus on proportional rather than rank-based changes.
To understand the rate of infrastructure development, we computed the normalized change in pavedness between 2020 and 2024. This change was standardized by each country's remaining paving potential, utilizing the formula:

\begin{equation}
\Delta P_{\text{norm}} = \frac{P_{2024} - P_{2020}}{1 - P_{2020}} \times 100
\label{eq:pavedness_change} % Optional: add a label for referencing
\end{equation}

where $P_{2020}$ and $P_{2024}$ represent the shares of paved roads in 2020 and 2024, respectively. This normalization ensures that the rate of change is expressed relative to each country's unpaved share at the baseline, thereby facilitating meaningful cross-country comparisons irrespective of initial infrastructure levels. To isolate the direct relationship between SHDI and relative improvements in road infrastructure, and to control for the inherent dependency of normalized change measures on their baseline, we applied partial regression, removing the effect of initial pavedness. This approach provided a clearer interpretation of how human development corresponds to the pace of infrastructural progress.

\bibliography{main}

\begin{thebibliography}{10}
\urlstyle{rm}
\expandafter\ifx\csname url\endcsname\relax
  \def\url#1{\texttt{#1}}\fi
\expandafter\ifx\csname urlprefix\endcsname\relax\def\urlprefix{URL }\fi
\expandafter\ifx\csname doiprefix\endcsname\relax\def\doiprefix{DOI: }\fi
\providecommand{\bibinfo}[2]{#2}
\providecommand{\eprint}[2][]{\url{#2}}

\bibitem{worldbank_transport_2017}
\bibinfo{author}{{The World Bank}}.
\newblock \emph{\bibinfo{title}{The World Bank Group’s Transport Universal Access Practice : A Guidance Note}} (\bibinfo{publisher}{The World Bank}, \bibinfo{address}{Washington, DC}, \bibinfo{year}{2017}).

\bibitem{10.1093/jeea/jvab027}
\bibinfo{author}{Jedwab, R.} \& \bibinfo{author}{Storeygard, A.}
\newblock \bibinfo{journal}{\bibinfo{title}{The average and heterogeneous effects of transportation investments: Evidence from sub-saharan africa 1960–2010}}.
\newblock {\emph{\JournalTitle{Journal of the European Economic Association}}} \textbf{\bibinfo{volume}{20}}, \bibinfo{pages}{1--38}, \doiprefix\url{10.1093/jeea/jvab027} (\bibinfo{year}{2021}).
\newblock \eprint{https://academic.oup.com/jeea/article-pdf/20/1/1/42763270/jvab027.pdf}.

\bibitem{un_sdg9}
\bibinfo{author}{{United Nations}}.
\newblock \bibinfo{title}{Goal 9: Build resilient infrastructure, promote inclusive and sustainable industrialization and foster innovation}.
\newblock \bibinfo{howpublished}{\url{https://sdgs.un.org/goals/goal9}} (\bibinfo{year}{2015}).

\bibitem{Wenz_2020}
\bibinfo{author}{Wenz, L.}, \bibinfo{author}{Weddige, U.}, \bibinfo{author}{Jakob, M.} \& \bibinfo{author}{Steckel, J.~C.}
\newblock \bibinfo{journal}{\bibinfo{title}{Road to glory or highway to hell? global road access and climate change mitigation}}.
\newblock {\emph{\JournalTitle{Environmental Research Letters}}} \textbf{\bibinfo{volume}{15}}, \bibinfo{pages}{075010}, \doiprefix\url{10.1088/1748-9326/ab858d} (\bibinfo{year}{2020}).

\bibitem{african2014tracking}
\bibinfo{author}{{African Development Bank Group}}.
\newblock \bibinfo{title}{Tracking africa’s progress in figures} (\bibinfo{year}{2014}).
\newblock \bibinfo{note}{Accessed: 2024-10-08}.

\bibitem{Queiroz1992Roads}
\bibinfo{author}{Queiroz, C.} \& \bibinfo{author}{Gautam, S.}
\newblock \bibinfo{title}{Road infrastructure and economic development: Some diagnostic indicators}.
\newblock \bibinfo{type}{Policy Research Working Paper} \bibinfo{number}{921}, \bibinfo{institution}{World Bank} (\bibinfo{year}{1992}).

\bibitem{calderon_roads_2015}
\bibinfo{author}{Calderón, C.} \& \bibinfo{author}{Servén, L.}
\newblock \bibinfo{journal}{\bibinfo{title}{The effects of infrastructure development on growth and income distribution}}.
\newblock {\emph{\JournalTitle{Journal of Economic Surveys}}} \textbf{\bibinfo{volume}{29}}, \bibinfo{pages}{34--72} (\bibinfo{year}{2015}).

\bibitem{GEBRESILASSE2023103048}
\bibinfo{author}{Gebresilasse, M.}
\newblock \bibinfo{journal}{\bibinfo{title}{Rural roads, agricultural extension, and productivity}}.
\newblock {\emph{\JournalTitle{Journal of Development Economics}}} \textbf{\bibinfo{volume}{162}}, \bibinfo{pages}{103048}, \doiprefix\url{https://doi.org/10.1016/j.jdeveco.2023.103048} (\bibinfo{year}{2023}).

\bibitem{Koks_2023}
\bibinfo{author}{Koks, E.} \emph{et~al.}
\newblock \bibinfo{journal}{\bibinfo{title}{A global assessment of national road network vulnerability}}.
\newblock {\emph{\JournalTitle{Environmental Research: Infrastructure and Sustainability}}} \textbf{\bibinfo{volume}{3}}, \bibinfo{pages}{025008}, \doiprefix\url{10.1088/2634-4505/acd1aa} (\bibinfo{year}{2023}).

\bibitem{doi:10.1126/science.aaf7894}
\bibinfo{author}{Jean, N.} \emph{et~al.}
\newblock \bibinfo{journal}{\bibinfo{title}{Combining satellite imagery and machine learning to predict poverty}}.
\newblock {\emph{\JournalTitle{Science}}} \textbf{\bibinfo{volume}{353}}, \bibinfo{pages}{790--794}, \doiprefix\url{10.1126/science.aaf7894} (\bibinfo{year}{2016}).
\newblock \eprint{https://www.science.org/doi/pdf/10.1126/science.aaf7894}.

\bibitem{yeh2020economic}
\bibinfo{author}{Yeh, C.} \emph{et~al.}
\newblock \bibinfo{journal}{\bibinfo{title}{Using publicly available satellite imagery and deep learning to understand economic well-being in africa}}.
\newblock {\emph{\JournalTitle{Nature Communications}}} \textbf{\bibinfo{volume}{11}}, \bibinfo{pages}{2583}, \doiprefix\url{10.1038/s41467-020-16185-w} (\bibinfo{year}{2020}).

\bibitem{facebook_map_with_ai}
\bibinfo{author}{{Meta}}.
\newblock \bibinfo{title}{Map with ai}.
\newblock \bibinfo{howpublished}{\url{https://mapwith.ai}} (\bibinfo{year}{2020}).
\newblock \bibinfo{note}{Accessed: 2024-09-11}.

\bibitem{Microsoft_RoadDetections}
\bibinfo{author}{{Microsoft}}.
\newblock \bibinfo{title}{Global road detections from satellite imagery}.
\newblock \bibinfo{howpublished}{\url{https://github.com/microsoft/RoadDetections}} (\bibinfo{year}{2021}).

\bibitem{microsoft_building_footprints}
\bibinfo{author}{Microsoft}.
\newblock \bibinfo{title}{Us building footprints dataset}.
\newblock \bibinfo{howpublished}{\url{https://github.com/microsoft/USBuildingFootprints}} (\bibinfo{year}{2020}).
\newblock \bibinfo{note}{Accessed: 2024-09-11}.

\bibitem{google_open_buildings}
\bibinfo{author}{{Google Research}}.
\newblock \bibinfo{title}{Google open buildings dataset}.
\newblock \bibinfo{howpublished}{\url{https://sites.research.google/open-buildings/}} (\bibinfo{year}{2021}).
\newblock \bibinfo{note}{Accessed: 2024-09-11}.

\bibitem{haklay_quality_2010}
\bibinfo{author}{Haklay, M.}
\newblock \bibinfo{journal}{\bibinfo{title}{How good is volunteered geographical information? a comparative study of openstreetmap and ordnance survey datasets}}.
\newblock {\emph{\JournalTitle{Environment and Planning B: Planning and Design}}} \textbf{\bibinfo{volume}{37}}, \bibinfo{pages}{682--703} (\bibinfo{year}{2010}).

\bibitem{su9060997}
\bibinfo{author}{Mobasheri, A.}, \bibinfo{author}{Sun, Y.}, \bibinfo{author}{Loos, L.} \& \bibinfo{author}{Ali, A.~L.}
\newblock \bibinfo{journal}{\bibinfo{title}{Are crowdsourced datasets suitable for specialized routing services? case study of openstreetmap for routing of people with limited mobility}}.
\newblock {\emph{\JournalTitle{Sustainability}}} \textbf{\bibinfo{volume}{9}}, \doiprefix\url{10.3390/su9060997} (\bibinfo{year}{2017}).

\bibitem{doi:10.1139/geomat-2021-0012}
\bibinfo{author}{Moradi, M.}, \bibinfo{author}{Roche, S.} \& \bibinfo{author}{Mostafavi, M.~A.}
\newblock \bibinfo{journal}{\bibinfo{title}{Exploring five indicators for the quality of openstreetmap road networks: a case study of québec, canada}}.
\newblock {\emph{\JournalTitle{Geomatica}}} \textbf{\bibinfo{volume}{75}}, \bibinfo{pages}{178--208}, \doiprefix\url{10.1139/geomat-2021-0012} (\bibinfo{year}{2021}).

\bibitem{RANDHAWA2025362}
\bibinfo{author}{Randhawa, S.} \emph{et~al.}
\newblock \bibinfo{journal}{\bibinfo{title}{Paved or unpaved? a deep learning derived road surface global dataset from mapillary street-view imagery}}.
\newblock {\emph{\JournalTitle{ISPRS Journal of Photogrammetry and Remote Sensing}}} \textbf{\bibinfo{volume}{223}}, \bibinfo{pages}{362--374}, \doiprefix\url{https://doi.org/10.1016/j.isprsjprs.2025.02.020} (\bibinfo{year}{2025}).

\bibitem{zhu_deep_2017}
\bibinfo{author}{Zhu, X.~X.} \emph{et~al.}
\newblock \bibinfo{journal}{\bibinfo{title}{Deep learning in remote sensing: A comprehensive review and list of resources}}.
\newblock {\emph{\JournalTitle{IEEE Geoscience and Remote Sensing Magazine}}} \textbf{\bibinfo{volume}{5}}, \bibinfo{pages}{8--36} (\bibinfo{year}{2017}).

\bibitem{10.1145/3615900.3628772}
\bibinfo{author}{Randhawa, S.} \emph{et~al.}
\newblock \bibinfo{title}{Multiscale multifeature vision learning for scalable and efficient wastewater treatment plant detection using hi-res satellite imagery and osm}.
\newblock In \emph{\bibinfo{booktitle}{Proceedings of the 1st ACM SIGSPATIAL International Workshop on Advances in Urban-AI}}, UrbanAI '23, \bibinfo{pages}{10–21}, \doiprefix\url{10.1145/3615900.3628772} (\bibinfo{publisher}{Association for Computing Machinery}, \bibinfo{address}{New York, NY, USA}, \bibinfo{year}{2023}).

\bibitem{cheng2022maskedattentionmasktransformeruniversal}
\bibinfo{author}{Cheng, B.}, \bibinfo{author}{Misra, I.}, \bibinfo{author}{Schwing, A.~G.}, \bibinfo{author}{Kirillov, A.} \& \bibinfo{author}{Girdhar, R.}
\newblock \bibinfo{title}{Masked-attention mask transformer for universal image segmentation} (\bibinfo{year}{2022}).
\newblock \eprint{2112.01527}.

\bibitem{DBLP:journals/corr/AlbertKG17}
\bibinfo{author}{Albert, A.}, \bibinfo{author}{Kaur, J.} \& \bibinfo{author}{Gonz{\'{a}}lez, M.~C.}
\newblock \bibinfo{journal}{\bibinfo{title}{Using convolutional networks and satellite imagery to identify patterns in urban environments at a large scale}}.
\newblock {\emph{\JournalTitle{CoRR}}} \textbf{\bibinfo{volume}{abs/1704.02965}} (\bibinfo{year}{2017}).
\newblock \eprint{1704.02965}.

\bibitem{aleissaee2022transformersremotesensingsurvey}
\bibinfo{author}{Aleissaee, A.~A.} \emph{et~al.}
\newblock \bibinfo{title}{Transformers in remote sensing: A survey} (\bibinfo{year}{2022}).
\newblock \eprint{2209.01206}.

\bibitem{wang2022selfsupervisedlearningremotesensing}
\bibinfo{author}{Wang, Y.}, \bibinfo{author}{Albrecht, C.~M.}, \bibinfo{author}{Braham, N. A.~A.}, \bibinfo{author}{Mou, L.} \& \bibinfo{author}{Zhu, X.~X.}
\newblock \bibinfo{title}{Self-supervised learning in remote sensing: A review} (\bibinfo{year}{2022}).
\newblock \eprint{2206.13188}.

\bibitem{zhou_mapping_2024}
\bibinfo{author}{Zhou, Q.}, \bibinfo{author}{Liu, Z.} \& \bibinfo{author}{Huang, Z.}
\newblock \bibinfo{journal}{\bibinfo{title}{Mapping {Road} {Surface} {Type} of {Kenya} {Using} {OpenStreetMap} and {High}-resolution {Google} {Satellite} {Imagery}}}.
\newblock {\emph{\JournalTitle{Scientific Data}}} \textbf{\bibinfo{volume}{11}}, \bibinfo{pages}{331}, \doiprefix\url{10.1038/s41597-024-03158-7} (\bibinfo{year}{2024}).

\bibitem{rs15163985}
\bibinfo{author}{Workman, R.}, \bibinfo{author}{Wong, P.}, \bibinfo{author}{Wright, A.} \& \bibinfo{author}{Wang, Z.}
\newblock \bibinfo{journal}{\bibinfo{title}{Prediction of unpaved road conditions using high-resolution optical satellite imagery and machine learning}}.
\newblock {\emph{\JournalTitle{Remote Sens. (Basel)}}} \textbf{\bibinfo{volume}{15}}, \bibinfo{pages}{3985} (\bibinfo{year}{2023}).

\bibitem{doi:10.1126/science.abe8628}
\bibinfo{author}{Burke, M.}, \bibinfo{author}{Driscoll, A.}, \bibinfo{author}{Lobell, D.~B.} \& \bibinfo{author}{Ermon, S.}
\newblock \bibinfo{journal}{\bibinfo{title}{Using satellite imagery to understand and promote sustainable development}}.
\newblock {\emph{\JournalTitle{Science}}} \textbf{\bibinfo{volume}{371}}, \bibinfo{pages}{eabe8628}, \doiprefix\url{10.1126/science.abe8628} (\bibinfo{year}{2021}).
\newblock \eprint{https://www.science.org/doi/pdf/10.1126/science.abe8628}.

\bibitem{planet}
\bibinfo{author}{PBC, P.~L.}
\newblock \bibinfo{title}{Planet application program interface: In space for life on earth} (\bibinfo{year}{2025}).

\bibitem{MRH}
\bibinfo{author}{of~Roads \&~Highways, M.}
\newblock \bibinfo{title}{A ministry of the republic of ghana}.
\newblock \bibinfo{howpublished}{https://mrh.gov.gh/?b=63259658875}.
\newblock \bibinfo{note}{Accessed:2025-09-11}.

\bibitem{GHA}
\bibinfo{author}{Authority, G.~H.}
\newblock \bibinfo{title}{State interests and governance authority,}.
\newblock \bibinfo{howpublished}{https://siga.gov.gh/entity/ghana-highway-authority/}.
\newblock \bibinfo{note}{Accessed:2025-09-11}.

\bibitem{DUR}
\bibinfo{author}{of~Urban~Roads, D.}
\newblock \bibinfo{title}{Department of urban roads, ghana}.
\newblock \bibinfo{howpublished}{https://dur.gov.gh/functions/}.
\newblock \bibinfo{note}{Accessed:2025-09-11}.

\bibitem{DFR}
\bibinfo{author}{of~Feeder~Roads, D.}
\newblock \bibinfo{title}{Functions-department of feeder roads, ghana}.
\newblock \bibinfo{howpublished}{\url{https://dfr.gov.gh/dfr-ftp/about\%20us.html}}.
\newblock \bibinfo{note}{Accessed: 2025-09-11}.

\bibitem{BILL}
\bibinfo{author}{{Parliament of Ghana}}.
\newblock \bibinfo{title}{National roads authority bill, 2023}.
\newblock \bibinfo{howpublished}{\url{http://www.commonlii.org/gh/legis/bill/nrab2023289.pdf}} (\bibinfo{year}{2023}).
\newblock \bibinfo{note}{Accessed: 2025-09-11}.

\bibitem{cross_road}
\bibinfo{author}{Brilé~Anderson, V. S.~A.}
\newblock \bibinfo{title}{Accra at a crossroads: Building a transport system for all}.
\newblock \bibinfo{howpublished}{https://www.oecd.org/en/blogs/2025/06/accra-at-a-crossroads-building-a-transport-system-for-all.html} (\bibinfo{year}{2025}).
\newblock \bibinfo{note}{Accessed:2025-09-11}.

\bibitem{YOR}
\bibinfo{author}{3News.com}.
\newblock \bibinfo{title}{Akufo-addo declares year of roads}.
\newblock \bibinfo{howpublished}{https://www.modernghana.com/news/1129919/akufo-addo-declares-year-of-roads.html}.
\newblock \bibinfo{note}{Accessed:2025-09-11}.

\bibitem{Gorgulu2023Infrastructure}
\bibinfo{author}{Gorgulu, N.}, \bibinfo{author}{Foster, V.}, \bibinfo{author}{Jain, D.}, \bibinfo{author}{Straub, S.} \& \bibinfo{author}{Vagliasindi, M.}
\newblock \bibinfo{title}{The impact of infrastructure on development outcomes: A meta-analysis}.
\newblock \bibinfo{type}{Policy Research Working Paper} \bibinfo{number}{10350}, \bibinfo{institution}{World Bank} (\bibinfo{year}{2023}).
\newblock \bibinfo{note}{License: CC BY-NC 3.0 IGO}.

\bibitem{Roberts_etal_2006}
\bibinfo{author}{Roberts, P.}, \bibinfo{author}{KC, S.} \& \bibinfo{author}{Rastogi, C.}
\newblock \emph{\bibinfo{title}{Measuring Rural Access: A Guide to Good Practice}} (\bibinfo{publisher}{The World Bank}, \bibinfo{address}{Washington, DC}, \bibinfo{year}{2006}).

\bibitem{sun2023ruralaccessindexglobal}
\bibinfo{author}{Sun, Q.}, \bibinfo{author}{Li, W.} \& \bibinfo{author}{Zhou, Q.}
\newblock \bibinfo{title}{Rural access index: A global study} (\bibinfo{year}{2023}).
\newblock \eprint{2309.00505}.

\bibitem{iimi2016ruralaccess}
\bibinfo{author}{Iimi, A.}, \bibinfo{author}{You, L.} \& \bibinfo{author}{Wood-Sichra, U.}
\newblock \bibinfo{title}{New rural access index: Main determinants and correlation to poverty}.
\newblock \bibinfo{type}{Policy Research Working Paper} \bibinfo{number}{7876}, \bibinfo{institution}{World Bank}, \bibinfo{address}{Washington, D.C.} (\bibinfo{year}{2016}).

\bibitem{Sun_etal_2023}
\bibinfo{author}{Sun, Y.~e.}
\newblock \bibinfo{journal}{\bibinfo{title}{Global and regional patterns of rural travel and accessibility}}.
\newblock {\emph{\JournalTitle{Nature Communications}}} \textbf{\bibinfo{volume}{14}}, \bibinfo{pages}{3641} (\bibinfo{year}{2023}).

\bibitem{Yang2024}
\bibinfo{author}{Yang, S.} \& \bibinfo{author}{Jin, Z.}
\newblock \bibinfo{journal}{\bibinfo{title}{Impact of road network expansion on landscape ecological risk and soil erosion sensitivity on the luochuan tableland of the chinese loess plateau}}.
\newblock {\emph{\JournalTitle{Regional Environmental Change}}} \textbf{\bibinfo{volume}{24}}, \doiprefix\url{10.1007/s10113-024-02202-x} (\bibinfo{year}{2024}).

\bibitem{ESA2021}
\bibinfo{author}{Coffin, A.~W.}
\newblock \bibinfo{title}{Rural roads: A construction and maintenance guide for protecting the environment} (\bibinfo{year}{2021}).
\newblock \bibinfo{note}{Accessed: 2024-10-04}.

\bibitem{herfort_spatio-temporal_2023}
\bibinfo{author}{Herfort, B.}, \bibinfo{author}{Lautenbach, S.}, \bibinfo{author}{Porto~de Albuquerque, J.}, \bibinfo{author}{Anderson, J.} \& \bibinfo{author}{Zipf, A.}
\newblock \bibinfo{journal}{\bibinfo{title}{A spatio-temporal analysis investigating completeness and inequalities of global urban building data in {OpenStreetMap}}}.
\newblock {\emph{\JournalTitle{Nature Communications}}} \textbf{\bibinfo{volume}{14}}, \bibinfo{pages}{3985}, \doiprefix\url{10.1038/s41467-023-39698-6} (\bibinfo{year}{2023}).

\bibitem{OSMNX}
\bibinfo{author}{Boeing, G.}
\newblock \bibinfo{journal}{\bibinfo{title}{Modeling and analyzing urban networks and amenities with osmnx}}.
\newblock {\emph{\JournalTitle{Geographical Analysis}}} \textbf{\bibinfo{volume}{57}}, \bibinfo{pages}{567--577}, \doiprefix\url{https://doi.org/10.1111/gean.70009} (\bibinfo{year}{2025}).
\newblock \eprint{https://onlinelibrary.wiley.com/doi/pdf/10.1111/gean.70009}.

\bibitem{hartigan1979algorithm}
\bibinfo{author}{Hartigan, J.~A.} \& \bibinfo{author}{Wong, M.~A.}
\newblock \bibinfo{journal}{\bibinfo{title}{Algorithm as 136: A k-means clustering algorithm}}.
\newblock {\emph{\JournalTitle{Journal of the royal statistical society. series c (applied statistics)}}} \textbf{\bibinfo{volume}{28}}, \bibinfo{pages}{100--108} (\bibinfo{year}{1979}).

\bibitem{dijkstra_1959}
\bibinfo{author}{Dijkstra, E.~W.}
\newblock \bibinfo{journal}{\bibinfo{title}{A note on two problems in connexion with graphs}}.
\newblock {\emph{\JournalTitle{Numerische Mathematik}}} \textbf{\bibinfo{volume}{1}}, \bibinfo{pages}{269--271}, \doiprefix\url{10.1007/BF01386390} (\bibinfo{year}{1959}).

\end{thebibliography}

\section*{Data Availability}

The global road surface dataset generated and analyzed in this study is openly available at the \textit{Humanitarian Data Exchange (HDX)} under a Creative Commons Attribution--NonCommercial 4.0 (CC BY-NC 4.0) license at \url{https://doi.org/10.xxxx/link-to-dataset}. The dataset includes all attributes detailed in Supplementary Table~\ref{S-tab:dataset_attributes}. 

Additional datasets used in this study are publicly available from the sources cited in the main text. The PlanetScope satellite imagery used for model training and validation was obtained under a research and education license agreement with Planet Labs PBC. Due to these licensing conditions, the Planet imagery itself cannot be redistributed.

\section*{Definition of Arterial Road Network}
\phantomsection
\label{sec:def_arterial}
In this study, we define the global arterial road network as comprising OpenStreetMap (OSM) \texttt{highway} tags classified as \texttt{motorway}, \texttt{trunk}, \texttt{primary}, and \texttt{secondary}, including their associated link types (e.g., \texttt{motorway\_link}).For more details see Supplementary Table \ref{S-tab:osm-highway-plain}. These classes were selected based on their functional role in enabling long-distance, high-capacity transportation and their critical importance for national and regional connectivity.For more details see Supplementary Table \ref{S-tab:osm-highway-plain}.

\section*{Acknowledgements}

We are grateful to Alec Schulze-Eckel for his expert insights on classifying road widths for humanitarian applications. We also thank Dr. Michael Auer for his initial work on big data analytics, Eren Aygün or processing the OSM road data, which provided the starting points for this dataset generation, Dr. Maciej Adamiak for helpful feedback on AI Modeling, Danille Gatland for her support in executing the 2020 deep learning pipelines and Till Frankenbach for valuable assistance with data visualization. We acknowledge the additional proofreading support provided by Prof. Supratik Guha and Lisa Shkredova.
The authors gratefully acknowledge support for the High-Performance Computing Infrastructure by the state of Baden-Württemberg through bwHPC and the German Research Foundation (DFG) through grant INST 35/1597-1 FUGG.
Moreover, the authors acknowledge the data storage service SDS@hd supported by the Ministry of Science, Research and the Arts Baden-Württemberg (MWK) and the German Research Foundation (DFG) through grant INST 35/1503-1 FUGG. 
We would like to extend our heartfelt thanks to the Klaus-Tschira Stiftung for their financial support.

\section*{Author contributions statement}

S.R. conceived and led the overall project and the overall analysis, and wrote the manuscript. G.R. finalized the global dataset, designed and developed the entire deep learning pipeline, including the big data acquisition, model training, validation, post-processing pipelines and data analysis. C.L. conducted the spatio-temporal analysis of the global results and generated the final maps. F.A. co-analyzed the results and insights for the Ghana and Pakistan case studies and co-wrote the corresponding sections. B.H., S.L., and A.Z. provided conceptual guidance through discussions and reviewed the manuscript. A.Z. also provided overall project management support. D.K. executed the deep learning pipeline to generate the results for the year 2020. O.O. created the labeled dataset and assisted with the human-the-in-the-loop validation. All authors reviewed and approved the final manuscript.

%\section*{Additional information}

%To include, in this order: \textbf{Accession codes} (where applicable); \textbf{Competing interests} (mandatory statement). 

%The corresponding author is responsible for submitting a \href{http://www.nature.com/srep/policies/index.html#competing}{competing interests statement} on behalf of all authors of the paper. This statement must be included in the submitted article file.

\section*{Supplementary Information}

The supplementary information is not included in this arXiv preprint due to size restrictions. It will be provided in the journal submission and can be made available upon request.

\end{document}